\documentclass[a4paper,11pt,fleqn]{cas-sc}

\usepackage[numbers,sort&compress]{natbib}

\usepackage{booktabs}
\usepackage{subfig}
\usepackage{setspace}   
\usepackage{amsmath}

\usepackage[section]{placeins}

\begin{document}
\let\WriteBookmarks\relax
\def\floatpagepagefraction{1}
\def\textpagefraction{.001}
\shorttitle{Wavelet transform analytics for UAV detection }
\shortauthors{OO Medaiyese et~al.}


\title [mode = title]{Wavelet Transform Analytics for RF-Based UAV Detection and
Identification System Using Machine Learning}

\tnotemark[1]

\tnotetext[1]{This work has been supported in part by NASA under the Federal Award ID number NNX17AJ94A, and by NSF CNS-1939334 Aerial Experimentation Research Platform for Advanced Wireless (AERPAW) project that supported the experiments at NC State.}

\author[1]{Olusiji Medaiyese}
\ead{o0meda01@louisville.edu}

\address[1]{Department of Computer Science and Engineering, University of Louisville,  Louisville, Kentucky,  40292, USA}

\author[2]{~Martins Ezuma}
\ead{mcezuma@ncsu.edu}

\author[1]{~Adrian Lauf}
\ead{aplauf01@louisville.edu}

\address[2]{Department of Electrical Engineering
North Carolina State University
Raleigh, North Carolina, 27606, USA}

\author[2]
{~Ismail Guvenc}
\ead{iguvenc@ncsu.edu}

\cortext[cor1]{Corresponding author}

\begin{abstract}
In this work, we performed a thorough comparative analysis on a radio frequency (RF) based drone detection and identification system (DDI) under wireless interference, such as WiFi and Bluetooth, by using machine learning algorithms, and a pre-trained convolutional neural network-based algorithm called SqueezeNet, as classifiers.  In RF signal fingerprinting research, the transient and steady state of the signals can be used to extract a unique signature from an RF signal. By exploiting the RF control signals from unmanned aerial vehicles (UAVs) for DDI, we considered each state of the signals separately for feature extraction and compared the pros and cons for drone detection and identification.  Using various categories of wavelet transforms (discrete wavelet transform, continuous wavelet transform, and wavelet scattering transform) for extracting features from the signals, we built different models using these features.  We studied the performance of these models under different signal to noise ratio (SNR) levels. By using the wavelet scattering transform to extract signatures (scattergrams) from the steady state of the RF signals at 30 dB SNR, and using these scattergrams to train SqueezeNet, we achieved an accuracy of 98.9$\%$ at 10 dB SNR.
\end{abstract}

\begin{graphicalabstract}

\end{graphicalabstract}

\begin{highlights}
\item To detect the presence of radio-controlled UAVs in an environment by exploiting the RF signal emanating from the UAV-flight controller communication under wireless interference (i.e., WiFi and Bluetooth).
\item To explore the possibility of extracting RF fingerprints from the transient and steady state of the RF signals for detection and identification of UAVs.
\item To utilize wavelet transform analytics (i.e., continuous wavelet transform and wavelet scattering transform) for the feature extraction where both coefficients and image-based signature are generated for training machine learning algorithms and convolutional neural network.
\item To evaluate the performance of trained models under varying signal to noise ratio.
\end{highlights}


\maketitle

\doublespacing

\section{Introduction}
Every new technology comes with both positive and negative impacts to our society. The unmanned aerial vehicles (UAVs) (commonly known as drone) technology is drastically evolving our world as its application is significantly broad. The immensity of UAV applications in different domains such as healthcare, logistics, remote sensing \cite{alsalam2017autonomous}, data acquisition \cite{banaszek2017application}, precision agriculture \cite{alsalam2017autonomous}, and environmental and disaster management \cite{coveney2017lightweight} has led to the birth of new business opportunities. UAVs are no longer restricted to military usage, as civilians are rapidly adopting it for both commercial and noncommercial use. Over 1.7 million UAVs are currently registered in the United States of America and more than 70 percent of the registered UAVs are used for recreational activities while the remaining percentage is used for commercial purposes \cite{uasbythenumbers_2020}.

 The growing use of UAVs is raising both security and privacy concerns. UAVs are known to have been used for cybercrimes, terrorism, drug smuggling, invading privacy, and other malicious intents \cite{rattledrone2016, smuggler_drone2018}. Aside from these malicious acts, there are few strict regulations on who can own or purchase a UAV. So, airspace contravention is another posing challenge of UAVs. For example, a hobbyist without the knowledge of airspace regulations can fly a UAV in restricted airspace and is therefore violating airspace regulations. Every month, the Federal Aviation Administration (FAA) receives over 100 cases or incidents of UAVs sighted in an unauthorized area in the United States \cite{faa_uas_sight_reporting2020}. These sightings are from pilots, citizens, and law enforcement, which involve human efforts as UAV detection mechanisms are still not prevalent.

UAV manufacturers have started enabling UAVs to have the geofencing capabilities. Geofencing is restricting the ability of a UAV to fly into or take off within some predefined areas (e.g., areas marked as No-Fly Zones) based on the GPS information. However, there is a need for a UAV detection system in geofencing-free areas that are sensitive to security and privacy. While there are radar or velocity-speed guns to detect when a driver is speeding on highways by law enforcement agents, no UAV detectors have wide-spread availability for airspace and infrastructure monitoring. There is no automatic system in place for law enforcement agencies to detect airspace contravention caused by UAVs. Solutions are still under development or at infant stages in development \cite{birch2015uas,doroftei2018qualitative,sturdivant2017systems}. Having this solution will enable law enforcement to tackle UAV crimes. Hence, the ever-increasing application of UAVs has made DDI research gain momentum in the domain of UAVs because of privacy and security issues \cite{brown2017pondering,nguyen2018cost}.

\begin{table*}
\centering
\caption{Comparison of  pros/cons for different UAV detection and tracking techniques.\label{Table1_comparison}}{%
\resizebox{\textwidth}{!}{%
\begin{tabular}{|p{2.8cm}|p{9 cm}|p{9 cm}|}

\hline
\bf{Detection~Technique}& \bf{Pros} & \bf{Cons} \\
\hline
 Audio (e.g.,~\cite{mezei2015drone,mezei2016drone,nijim2016drone,yue2018software})
 & It is cost-effective in implementation \cite{shi2018anti}, its operation is passive \cite{birch2015uas}.
 & Low detection accuracy, difficult to estimate the maximum detection range \cite{birch2015uas}, performance susceptible to ambient noise, high complexity in acquiring and maintaining an audio signature database, it cannot be used for UAVs with noise cancellation \cite{nguyen2018cost}.\\

\hline
Visual Imaging (e.g., \cite{saqib2017study, schumann2017deep, unlu2019deep}

& Commercial-off-the-shelf solutions that can be rapidly utilized, it can be used for detecting autonomous UAVs.

& Illumination of the environment or property under surveillance can degrade the performance, vulnerability to weather conditions, ineffectiveness for a crowded or cluttered area (i.e., line-of-sight (LOS) between the surveillance camera and the target is essential), a high-quality lens with an ultra-high-resolution may be needed to detect UAV at a long-range, it has limited coverage for a large environment as a camera can only focus on one direction per unit time. \\
\hline

Thermal~sensing (e.g., \cite{andravsi2017night})

& It is less susceptible to weather fluctuation and background clutter \cite{birch2015uas}
& UAVs have a low thermal signature, and it has limited coverage when a UAV is at non-line-of-sight (NLOS).\\
\hline

 Radar (e.g., \cite {drozdowicz201635, mendis2016deep})  & It can be used for detecting autonomously controlled UAVs, and it is not susceptible to weather fluctuation.

 & It induces interference on other RF bands, especially in a crowded environment \cite{nguyen2018cost}, it is not effective for UAV detection because the radar cross-section (RCS) depends on the absolute size of a flying object, making radar not effective, high cost of deployment, it requires line-of-sight to target for effective and efficient operation, and high powered radars are not recommended for areas with crowded human habitats because of their high active electromagnetic energy \cite{shi2018anti}. \\
 \hline
	RF  (e.g., \cite{zhao2018classification,zhou2018unmanned,ezuma2019micro,ezuma2019detection})
	
	& It is relatively inexpensive, that radio-controlled UAVs propagate RF signals, that its operation is stealthy, the size of the UAV does not affect its effectiveness, that it works for both line-of-sight and non-line-of-sight, that it is not dependent on protocol standardization \cite{zhao2018classification}, and that beyond using it for detecting the presence of a UAV, it can be exploited for the identification of operation modes of UAV \cite{al2019rf,alipour2019machine}.
	& It cannot be used to detect autonomously controlled UAVs.However, autonomously controlled UAVs are yet to reach maturity in the developmental process.It is not a trivial task to use RF-based detection because UAV radios operate at the industrial, scientific and medical (ISM) band and because there are several other devices (e.g., WiFi and Bluetooth) that operate at this same band, making it challenging to capture the UAV's RF signature.  \\
 \hline
\end{tabular}}}{}
\end{table*}

The three mitigating steps for curbing security or safety threatening-UAVs are detection, identification, and neutralization \cite{birch2015uas}. Detection involves using sensors to capture necessary information that shows some attributes of the presence of a UAV in a vicinity. Identification, which is also called classification, involves identifying the target based on the data provided at the detection stage. The neutralization phase relies on the output of the classification phase. If the classification phase identifies a target, then the alarm is raised with a counter-measure to bring down the UAV if necessary. Counter-measures may include jamming the RF communication between the UAV-flight controller, shooting down the UAV, and other methods \cite{birch2015uas}. It is important to note at this point that our paper focuses on detection and identification.

Several sensing techniques have been exploited in the literature for detection. These include audio, visual imaging, thermal sensing, RADAR, and radio frequency (RF) signature detection. For the audio-based technique, most UAVs generate sounds (i.e., a humming sound) when in operation. The sound waves propagated from a given UAV when in operation are adopted as the audio fingerprint or signature of the UAV. Visual imaging techniques involve the use of video surveillance cameras to monitor a restricted area. The rotor or motor of UAVs emits heat and the thermogram of this heat energy can be used as a thermal signature for the UAV. This approach is called the infrared thermography approach or thermal sensing. Radar is the most commonly used mechanism for detecting flying objects and is found at every airport around the world. A radar system uses a radio transmitter to propagate radio pulses toward an object so that the object can reflect the radio waves. The reflected waves are received by the receiver and the temporal properties of the reflected radio waves can be used to determine or identify the object.

 Most UAVs use a radio-controlled (RC) communication system. The radio pulses of the communication between the UAV and its flight controller can be intercepted and collected as an RF signature for UAV detection. Using these RF signals as signatures is based on the premise that each UAV-flight controller communication has unique features that are not necessarily based on the modulation types or propagating frequencies but may be a result of imperfection in the communication circuitries in the devices.

 Table \ref{Table1_comparison} provides the comparison of these detection techniques. While there are advantages and disadvantages to each of the techniques, we adopted the RF approach because its operation is stealthy, that it can be exploited to determine the operation mode of UAVs, and  that it can detect UAVs that are in non-line-of-sight. The major challenge of RF detection approach is that other wireless devices (e.g., WiFi and Bluetooth) operate at the same frequency band as that of UAV communication band. However, we propose different approaches that reliably classify WiFi, Bluetooth, and UAV signals in this work by using wavelet transform techniques for feature extraction.
 The contributions of this work are summarized as follows:\looseness=-1

\begin{enumerate}

\item We propose four different methods for extracting features or signatures from RF signals by using continuous wavelet transforms and wavelet time scattering transforms. To the best of our knowledge, this is the first time the four methods proposed in this work have been used for DDI systems or RF device identification.

\item We compared and contrast the effectiveness of using the image-based feature (scalogram and scattergram) over the coefficients (wavelet and scattering coefficients) based feature and the impact of a linear reduction in the dimensionality of the coefficients by using principal component analysis (PCA).

\item While it is very common to use the transient state of RF signal for signature extraction in literature, this paper investigates the performance of adopting either transient or steady states of RF signal for feature extraction under the proposed four methods for DDI systems. We demonstrated that the steady state of an RF signal contained unique attributes, and it is resilient to low SNR when scattergram is extracted.

\item We introduce the concept of transfer learning for UAV identification by using a publicly-available pre-trained network called SqueezeNet because it is an optimized neural network and a portable model in terms of size in memory.

\end{enumerate}

The remainder of the paper is organized as follows.  Section~\ref{two} provides a brief background and overview of the related work. Section~\ref{three} describes the system modeling for an RF-based UAV detection system. Section~\ref{four} introduces the experimental setup and data capturing steps. Section~\ref{five} provides the feature extraction methods proposed. Section~\ref{six} describes the UAV classification algorithm.  In Section~\ref{seven}, the performance evaluation and results are discussed. We provide conclusions and future work in Section~\ref{eight}.

\section{Background and Related Work} \label{two}

\begin{figure}
\center{\includegraphics[trim=0.1cm 0cm 0.1cm 0.1cm, clip,scale=0.33]{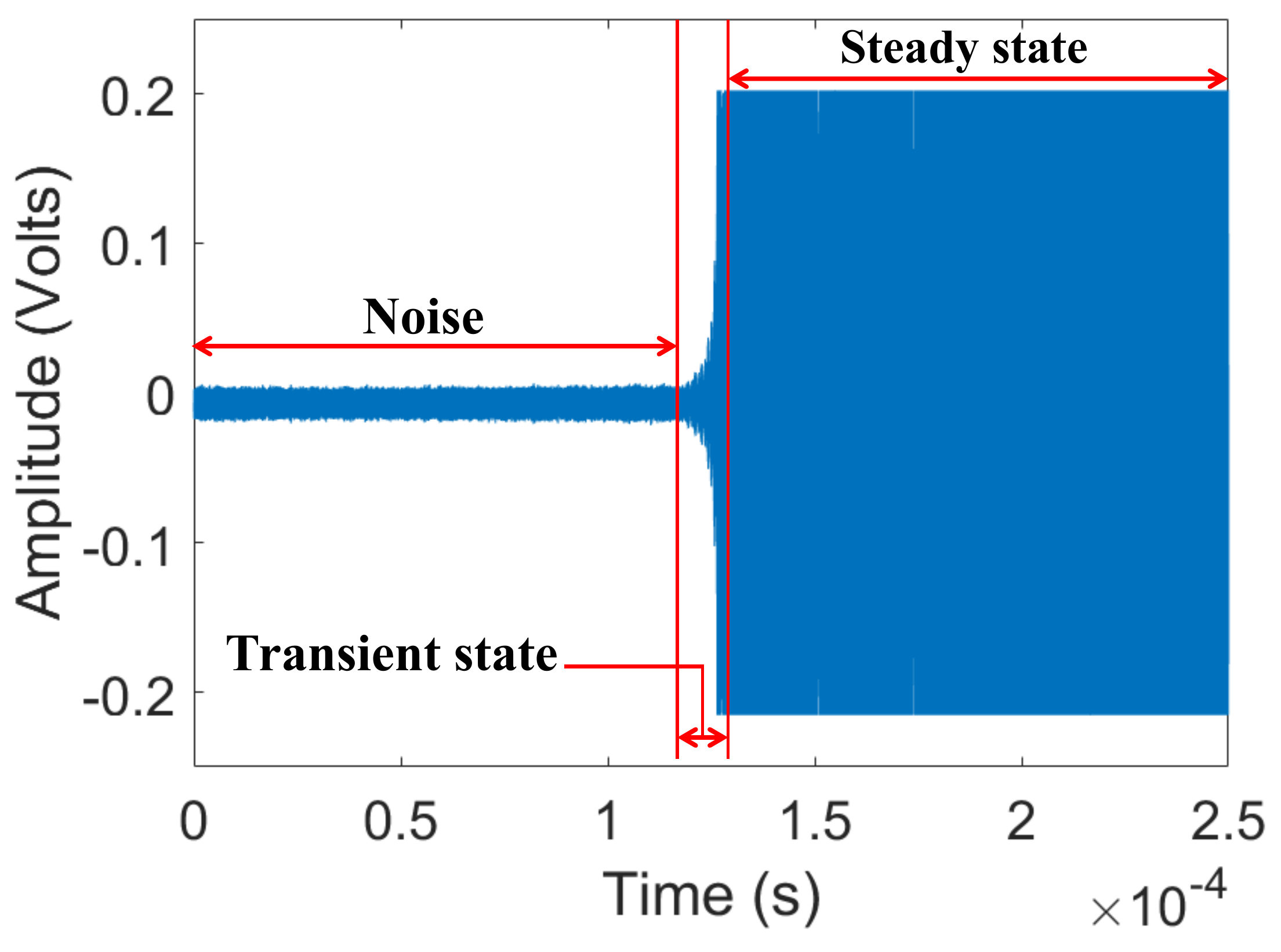}}
\caption{RF signal from DJI Phantom 4 controller with various parts (i.e., transient and steady state) of the signal labeled. }
\label{Fig:signal_label}
 \vspace{-5mm}
\end{figure}

Fig.~\ref{Fig:signal_label} shows the signature of a UAV controller (i.e., DJI Phantom 4 controller). The signature of an RF device is embedded in its RF signal. The uniqueness of signature is inherently attributed to the electronic circuitries (e.g., diode, capacitors, resistors, filters, PCB, soldering, etc.) of the device \cite{kennedy2008radio}. Direct usage of a signal can inhibit the computational and memory performance of the detection and identification system \cite{klein2009application}. Hence, the extraction of a signature is an imperative step that must be done carefully. The properties of RF fingerprint extraction include \cite{klein2009application}:
\begin{itemize}
  \item Minimal feature set to enable efficient time and space complexity
  \item Intra-device repeatability and stability
  \item Inter-device uniqueness
\end{itemize}
Selecting the representative features of the signal and measuring the discrimination between two signals are the challenges of signal classification especially when the signals are propagating at the same frequency \cite{fulcher2014highly} \cite{wang2013experimental}.

 In RF fingerprint-based DDI, the first step is the detection of a signal, followed by extracting unique features (i.e., fingerprint) from the detected signal. There are two parts of a detected signal, which are the transient and steady state. Fig.~\ref{Fig:signal_label} shows a typical example of a captured RF signal from a UAV controller. The part of the signal is explicitly labeled in the figure. The transient state is short-lived, and it occurs during power on and off of devices\cite{hall2003detection}. On the other hand, a steady state is a reliable detection source. It has a long period compared to the transient state and it is the state that occurs between the interval of the start and end of transient state\cite{kennedy2008radio}.  In the literature, the transient or steady state of an RF signal has been exploited to extract characteristics such as instantaneous amplitude, phase, frequency, and energy envelope as representative of the RF signature.

A transient signal has a unique amplitude variation that makes device distinguishable\cite{kennedy2008radio}. It has been established that the transient part has a good performance in device identification in literature. The steady state signal may contain unmodulated or modulated data. The variation in this data can make the steady state of an RF signal lack unique properties that can be used as a signature. However, in  \cite{klein2009application}, the authors used a non-transient state of orthogonal frequency-division multiplexing (OFDM)-based 802.11a signals for device identification. Similarly, the authors in \cite{kennedy2008radio} used the steady state spectral features for the identification of a radio transmitter.

Signal classification can be done either by instance-based classification or feature-based classification \cite{fulcher2014highly}. The instance-based classification computes the distances between the time series. Conversely, the feature-based classification compares sets of extracted features from the signals. In this paper, we use feature-based approaches for UAV detection and identification under inference like Bluetooth, and WiFi by exploiting wavelet analytics.

Table \ref{Table_ralated_work} provides the summary of some related work on the RF detection technique, comparison based on the feature extraction method, the part of the RF signal used for signature, performance evaluation (i.e., based on accuracy and inference time) and whether wireless inferences are considered.

In \cite{zhao2018classification}, an Auxiliary Classifier Wasserstein Generative Adversarial Networks (AC-WGANs) based model for UAV detection was proposed, which uses RF signatures from UAV radios as the input feature. Generative adversarial networks (GAN), which is commonly used for image generation and processing was adapted for detection and multi-classification of UAVs. This is done by exploiting and improving the discriminator model of the GAN. The amplitude envelope is used to reduce the length of the original signal and principal component analysis (PCA) is further employed to reduce the dimensionality of the RF signature for feature extraction. An accuracy of 95\% was obtained using AC-WGANs. However, the authors do not specify which part of the signal was used.

In \cite{zhou2018unmanned}, the channel state information (CSI) from the UAV RF channel was used to detect the presence of a UAV in an environment by examining the effect of mobility, spatiality, and vibration in the CSI. A rule-based algorithm was used for detection based on the metrics of the three effects. The accuracy of the rule-based model was 86.6\%

In \cite {al2019rf}, discrete fourier transform (DFT) was applied to UAV RF signals extract the frequency components of the signals, and the magnitude of the frequency components was used to train a deep neural network (DNN) model. An accuracy of 84.5\% was obtained when detecting and identifying the type of UAV. The authors did not consider other ISM devices that operate at the same frequency band (i.e., 2.4 GHz) as UAV-flight controller communication.

Similarly, the authors in \cite{alipour2019machine} used DFT to transform UAV signals to the frequency domain and  2048 frequency bins were obtained. A multi-channel 1-D convolutional neural network (CNN) was employed to extract features from the bins. The feature representations were then used to train a fully connected neural network. An accuracy of 94.6\% was achieved when identifying the type of UAVs. The effect of SNR on model performance and inference time was not considered or evaluated.

In \cite{ezuma2019micro,ezuma2019detection}, the authors proposed a k-nearest neighbours (\textit{k}NN) based model for detecting UAV using RF signatures (control signals) under the presence of WiFi and Bluetooth signals. Bandwidth resolution was used to identify UAV signals from both WiFi and Bluetooth. To classify the type of UAV, a two-level discrete wavelet transform (DWT) was used to decompose from the UAV signals, and statistical estimation (such as mean, shape factor, kurtosis, variance, and so on) of detailed coefficients was to train the \textit{k}NN model. An accuracy of 98.13\% was achieved.

\begin{table*}
\setlength{\tabcolsep}{1pt}
\centering

\caption{Related work on RF based UAV Detection and Identification.}

\label{Table_ralated_work}

\resizebox{\textwidth}{!}{%
\begin{tabular}{|c|c|c|c|c|c|c|c|}
\hline
 Literature & Feature extraction method& \multicolumn{2}{c|}{State of the signal}  & Algorithms & Accuracy & Interference & Inference time \\
 & & Transient & Steady &  & & &\\

 \hline
\cite{zhao2018classification}& PCA & N/A & N/A  & AC-WGANs & 95\%& None & No\\

\hline
\cite{zhou2018unmanned}& CSI& N/A & N/A&  rule-based algorithm & 86.6\%& WiFi & No\\

\hline

\cite{ezuma2019micro}& DWT, statistical features & Yes & No & ML algorithms &  96.3\%& None & No\\
 \hline

\cite{ezuma2019detection}& DWT, statistical features & Yes & No & ML algorithms &98.13\%& Bluetooth, WiFi& Yes\\
 \hline

\cite{al2019rf}& DFT & N/A & N/A  & DL & 84.5\%&None& No\\

\hline
\cite{allahham2020deep}& DFT and CNN & N/A & N/A  & DL & 94.6\%& None & No \\

\hline
\cite{ozturk2020rf}& spectrogram & Yes & N/A  & CNN & 99.5\%& Bluetooth, WiFi & No \\
\hline
{In this work} & CWT, WST & Yes & Yes &  ML and DL algorithms &98.9\%& Bluetooth, WiFi & Yes\\
\hline
\end{tabular}}
 \vspace{-2mm}
\end{table*}

\section{System Modeling for UAV Detection System}\label{three}

\begin{figure}
\center{\includegraphics[scale=0.5]{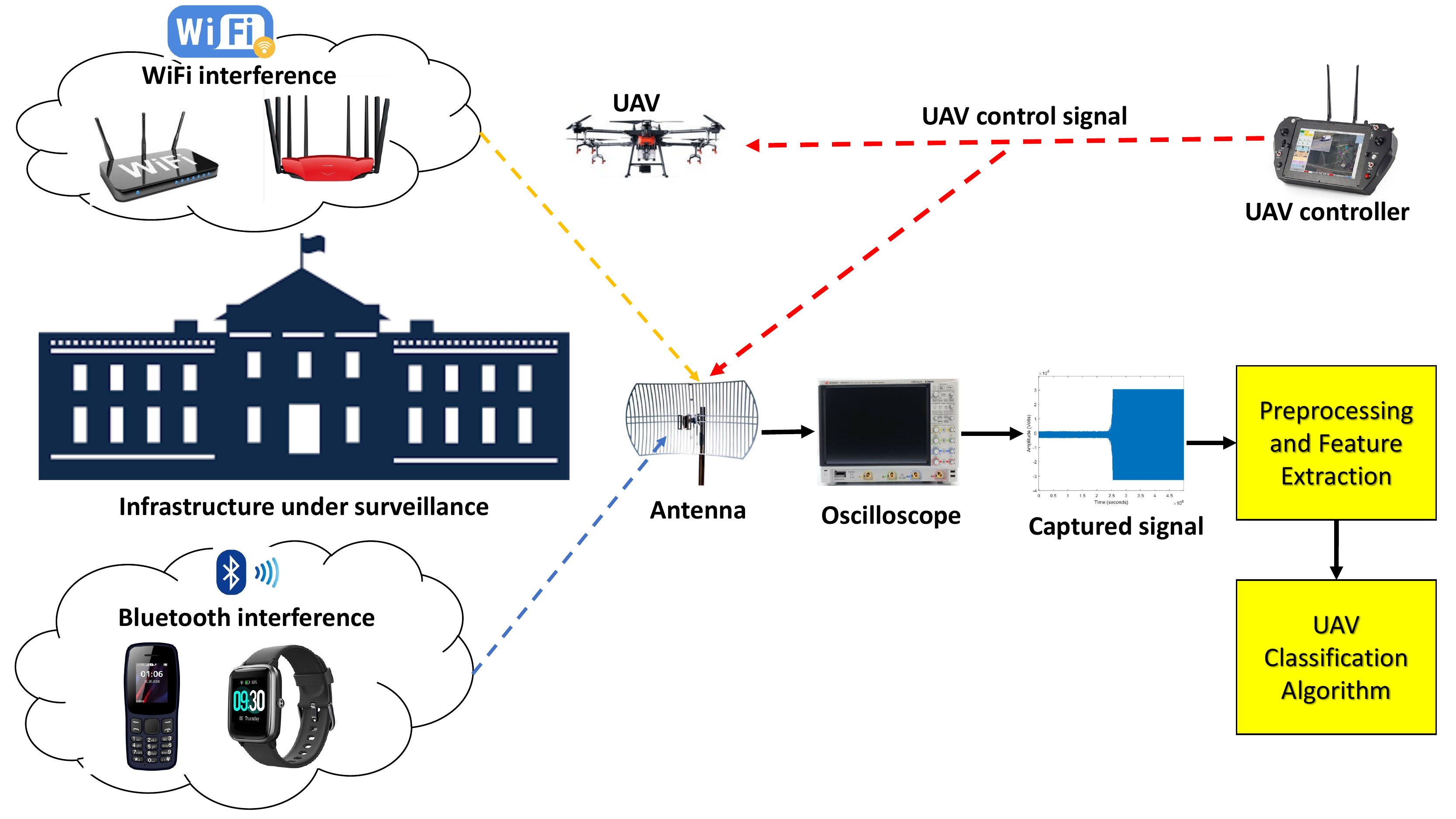}}
\caption{ RF-based UAV detection system for infrastructure surveillance under WiFi and Bluetooth interference.The UAV controller signals is exploited for the detection of UAV. UAV controller, WiFi, and Bluetooth devices operate at 2.4~GHz frequency band. An antenna is use to intercept RF signals from UAV controller, WiFi, and Bluetooth devices. The captured signal is processed and use for UAV detection and classification.}
\label{Fig:UAV_detection_scenario}
\end{figure}

Fig.~\ref{Fig:UAV_detection_scenario} shows the system modeling of an RF-based UAV detection system utilize for infrastructure surveillance. The system modeling gives an overview of our methodology. UAV controllers, WiFi, and Bluetooth devices operate at the same frequency band. So, a signal interceptor (antenna) and oscilloscope are used for signal capturing at 2.4 GHz in a stealth manner. The captured RF signals (i.e., from UAV controllers, WiFi, or Bluetooth devices) which serve as the signature to associated devices are preprocessed and use for extraction of features by exploiting wavelet transform analytics. The extracted features are used to train ML and CNN algorithms for UAV detection and classification.

\section{Experimental Setup and Data Capturing}\label{four}

\subsection{Data Capturing Step}
We captured RF signals from two Bluetooth devices (a smartphone and smart wristwatch), two WiFi routers, and six UAV controllers (four DJI UAVs, one BeeBeerun UAV, and one 3DR UAV). The operational frequency of all the devices is 2.4~GHz. Table ~\ref{UAV_catalogue} shows the catalog of the devices used for the experiment.

\begin{figure}
\center{\includegraphics[scale=0.45]{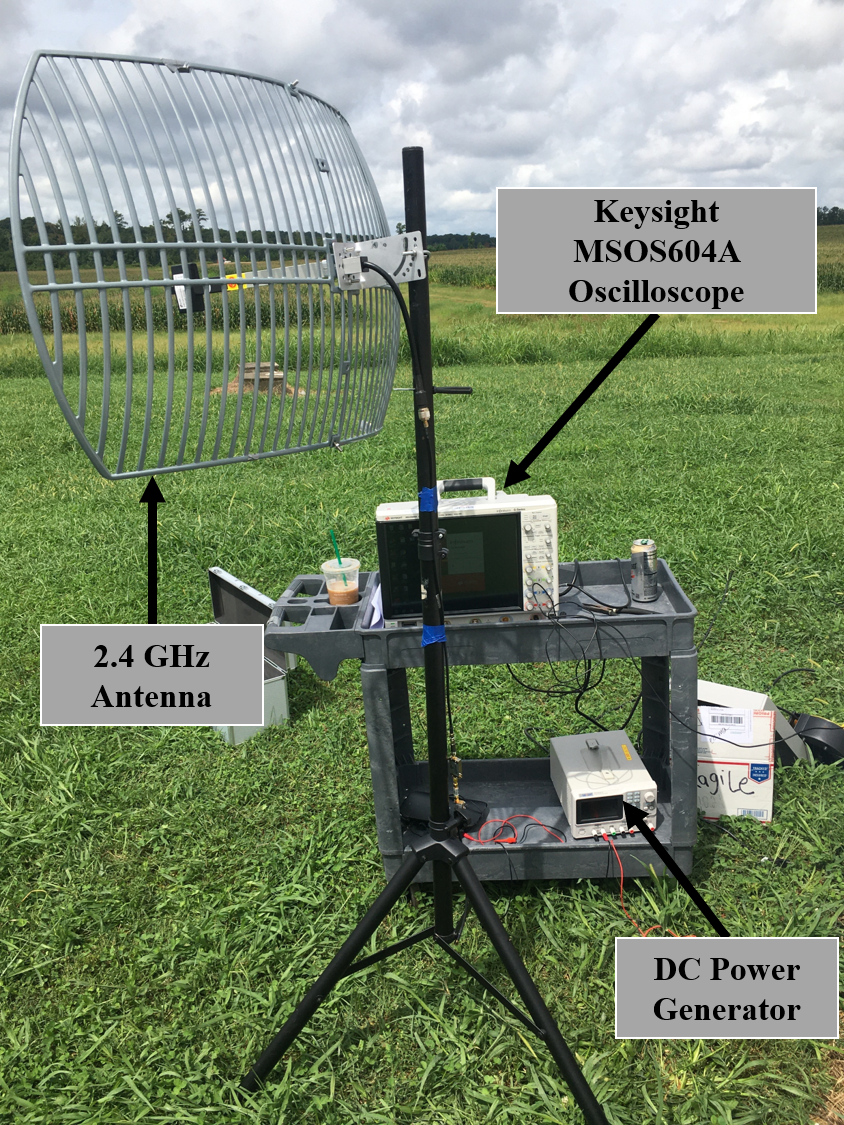}}
\caption{Experimental setup for capturing RF signals from UAV controllers, WiFi and Bluetooth devices in an outdoor setting. }
\label{Fig:detection_setup}
 \vspace{-5mm}
\end{figure}

The data was collected in an outdoor setting under a controlled environment. Fig.~\ref{Fig:detection_setup} shows the outdoor experimental setup. A 24 dBi 2.4 GHz grid parabolic antenna was used to intercept propagating RF signals from the RF devices (i.e., UAV controllers, Bluetooth, and WiFi devices). The intercepted signal goes through a 2.4~GHz bandpass filter which ensures that a frequency band at 2.4 GHz is acquired.  An RF low noise amplifier (LNA), FMAM63007, which operates from 2 GHz to 2.6~GHz with 30~dB gain is used to amplify the bandpass signal. A DC (direct current) generator is utilized to power the low noise amplifier attached to the bandpass filter. A 6 GHz bandwidth Keysight MSOS604A oscilloscope which has a sampling frequency of 20 GSa/s collects and stores the captured RF signals from the devices.

\begin{table}
\centering
\caption{ Catalog of RF devices used in the experiment for RF fingerprints acquisition. Under the UAV device, we use the UAV controllers from respective models.}
\label{UAV_catalogue}
\begin{tabular}{|c|c|c|}
\hline
Device & Make & Model\\
\hline
UAV &\multirow{4}{*}{\text{ DJI}} & Phantom 4  \\
&&Inspire  \\
&&Matrice 600  \\
&&Mavic Pro 1  \\\cline{2-3}

& Beebeerun & FPV RC drone mini quadcopter\\\cline{2-3}

& 3DR & Iris FS-TH9x\\
\hline
\multirow{2}{*}{\text{Bluetooth }}& Apple & Iphone 6S\\\cline{2-3}
& FitBit  &  Charge3 smartwatch\\
\hline
\multirow{2}{*}{\text{WiFI}} &Cisco & Linksys E3200\\\cline{2-3}
&TP-link & TL-WR940N\\

\hline
\end{tabular}
 \vspace{-3mm}
\end{table}

The oscilloscope has a threshold trigger for signal detection. We observed the background noise level in the environment from the oscilloscope and calibrated the threshold above the background noise level. In the presence of a signal, the energy level of the signal goes above the threshold and triggers the oscilloscope to capture and store the signal detected. In the absence of a signal, the oscilloscope does not capture data.  The captured data becomes the raw RF signals that are preprocessed for the classification algorithm.

Each captured signal consists of five million sampling points. For each device considered, 300 RF signals are collected at 30~dB SNR. This implies that 3000 RF signals are collected from the ten devices which take 20.1 GB of storage. We selected 200 RF signals from each device for the training set and the remaining 100 RF signals for the test set. So, 2000 RF signals are used for training purposes, and 1000 RF signals are utilized for testing.

\subsection{Pre-processing of Signal Using Wavelet Decomposition} \label{HWD_section}
\begin{figure}
\center{\includegraphics[scale=0.75]{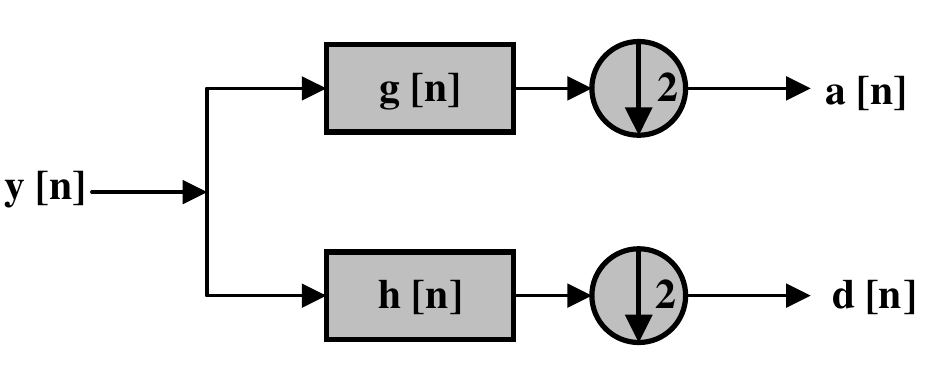}}
\caption{Single level Haar wavelet decomposition for raw RF signal prepossessing. The raw captured signal from RF devices is denoted as $y[n]$. The low pass and high pass filter are  $g[n]$ and $h[n]$ respectively. The approximate and detail coefficients are denoted as $a[n]$ and $d[n]$. Approximate coefficient,$a[n]$, is used for feature extraction.  }
\label{Fig:haar_decomposition}
 \vspace{-5mm}
\end{figure}

In selecting the beginning of a device's RF signal, we use a statistical changepoint detection algorithm in MATLAB to find abrupt changes in signal. The changepoint detector help us to determine the signal states (i.e., transient and steady states) and to prevent the noisy part of the raw signal from corrupting our RF signature. From here, the captured RF signal is then decomposed or preprocessed using a single level Haar wavelet decomposition (HWD) to improve the computational efficiency of extracting features. Fig.~\ref{Fig:haar_decomposition} shows the architecture of the HWD. The raw captured signal $y[n]$ is passed into two parallel-connected filters which are low pass filter ($g[n]$) and high pass filter ($h[n]$) respectively. The output of each filter is down-sampled. The output from down-sampling the outcome of the low pass filter, $a[n]$,  are called the approximation coefficients which represents the low-frequency component of the raw signal. Similarly, the outputs from down-sampling the results of the high pass filter,  $d[n]$, are called the detail coefficients which is the high-frequency component of the signal. Both the transient and steady states of the signal are acquired from the approximate coefficients, $a[n]$, of the decomposed signal. Features or RF fingerprints are acquired from these states for classification.

\section{Wavelet Analytics-based Feature Extraction}\label{five}
Time-frequency domain analysis is an essential step in extracting unique attributes (signatures) from a signal. The Fourier transform (FT) is the most applied mathematical function for analyzing the frequency content of a signal in signal processing. It essentially represents a signal, in the time domain, by series of sines and cosines in what is called the frequency domain, revealing several important features of the signal that were hidden or unknown in the time domain. While the FT shows the frequency of the sample, it does not give the frequency variation with time and it is of limited application particularly with signals of varying frequencies such as non-stationary signals. The Short-Time Fourier Transform (STFT) overcomes this limitation by sliding a window through the signal at a short interval/time, along the time axis, and performing FT on the data within that box.

The outcome of STFT is essentially a decomposition of the signal in the time domain into a time-frequency representation, which gives the frequency variation of the signal over time. The effectiveness of the STFT depends on the choice of window function used. The challenge with STFT is finding a suitable window size that balances time resolution with frequency resolution- choosing a window that gives higher frequency resolution gives a lower time resolution and vice versa.

The Wavelet Transform (WT) is one that seeks to mitigate the challenges of STFT. Unlike the STFT which slides a fixed window through the signal, the wavelet transform utilizes a variable window function \cite{mallat1999wavelet}.
The wavelet transform is the convolution of a base wavelet function with the signal under analysis or consideration \cite{adisson2002illustrated}.
Wavelet transforms enable the transformation of a signal in the time domain into a domain that will allow us to analytically examine hidden properties or features that describe the signal.

The WT of signal $f(t)$ is given as \cite{rioul1991wavelets}
\begin {equation} \label{eq:5}
w(s,\tau)=\frac {1}{\sqrt{s}}\int_{-\infty}^{\infty} f(t)  \psi( \frac{t-\tau}{s}) dt,
\end{equation}
where $s>0$ is scaling factor,  $\psi( \frac{t-\tau}{s})$ is the template function-based wavelet and $\tau$ is the time shifting factor.
The wavelet transforms takes a signal and decomposes it into a set of basis functions \cite{gao2010wavelets}. These basis functions are obtained by using a single template function (base wavelet) to perform both scaling and shifting operations along the time axis. Scaling is the process of stretching or shrinking the wavelet to match the feature of interest. The scaling factor, $s$, is used to categorize this process. The higher the $s$, the higher the stretch, and the lower the $s$ the higher the shrink.  The scaling factor is inversely proportional to the frequency of the signal \cite{gao2010wavelets}. The shrinking factor, $\tau$, is used to move the signal along the time axis.
 In wavelet transform, the similarities between the signal and a template function-based wavelet can be extracted during the decomposition process. These similarities are in a form of wavelet coefficients \cite{rioul1991wavelets} and they can represent underlying features or attributes in the signal.

\begin{figure}
\center{\includegraphics[scale=0.5]{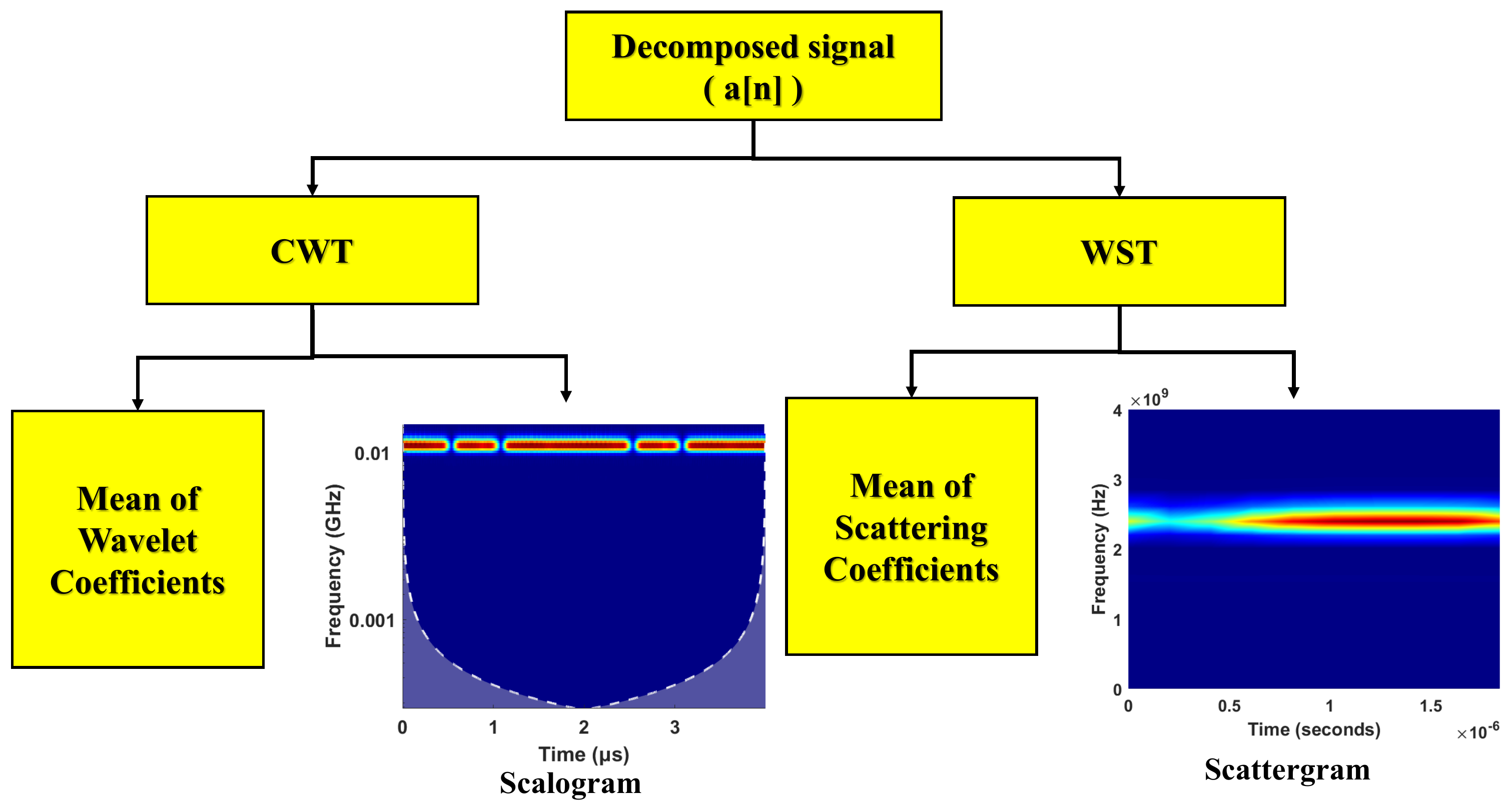}}
\caption{Divisions of the four proposed feature extraction methods based on CWT and WST. The decomposed signal, $a[n]$, is the approximation coefficient from the signal HWD.}
\label{Fig:feature_extraction_methods}
 \vspace{-5mm}
\end{figure}

We proposed four different methods that are based on wavelet transforms which are grouped into continuous wavelet transform (CWT) and wavelet scattering transform (WST). Fig.~\ref{Fig:feature_extraction_methods} shows the categories of the feature extraction methods. The approximate coefficients, $a[n]$, of the raw signal are used to extract signatures by using wavelet transforms. We compare and contrast the performance of these four feature extraction techniques.

\subsection{Feature Extraction based on Continuous Wavelet Transform}

Continuous wavelet transform (CWT) is defined as in (\ref{eq:5}). A derivative of STFT is the spectrogram, which is also used for analyzing signals in the time-frequency domain and it is the squared value of the STFT that provides the energy distribution in the signal in the time-frequency domain \cite{rioul1991wavelets}. Similarly, CWT provides a wavelet spectrogram which gives the distribution of energy in the signal in the time-scale domain. The wavelet spectrogram, also known as scalogram, is the squared modulus of the CWT or a plot of the energy density $E(s,\tau)$ \cite{rioul1991wavelets}.
\begin {equation} \label{eq:6}
E(s,\tau)=|w(s,\tau)|^2.
\end{equation}
The scalogram of a signal $x(t)$ can serve as the signature for the signal.
Morlet, Mexican hat, Gaussian, frequency B-Spline, harmonics, and Shannon wavelets are examples of base wavelets commonly used for CWT \cite{adisson2002illustrated,gao2010wavelets}.

The implementation of CWT by continuously varying the scale parameter, $s$, and translation parameter, $\tau$, introduces redundant information that may not be of value to the specific application \cite{gao2010wavelets}. To eliminate this redundancy, the scale and translation parameters are discretized leading to the discrete wavelet transform \cite{adisson2002illustrated,gao2010wavelets}

\begin{figure*}
\center{
\begin{subfloat}[]{\includegraphics[scale=0.35]{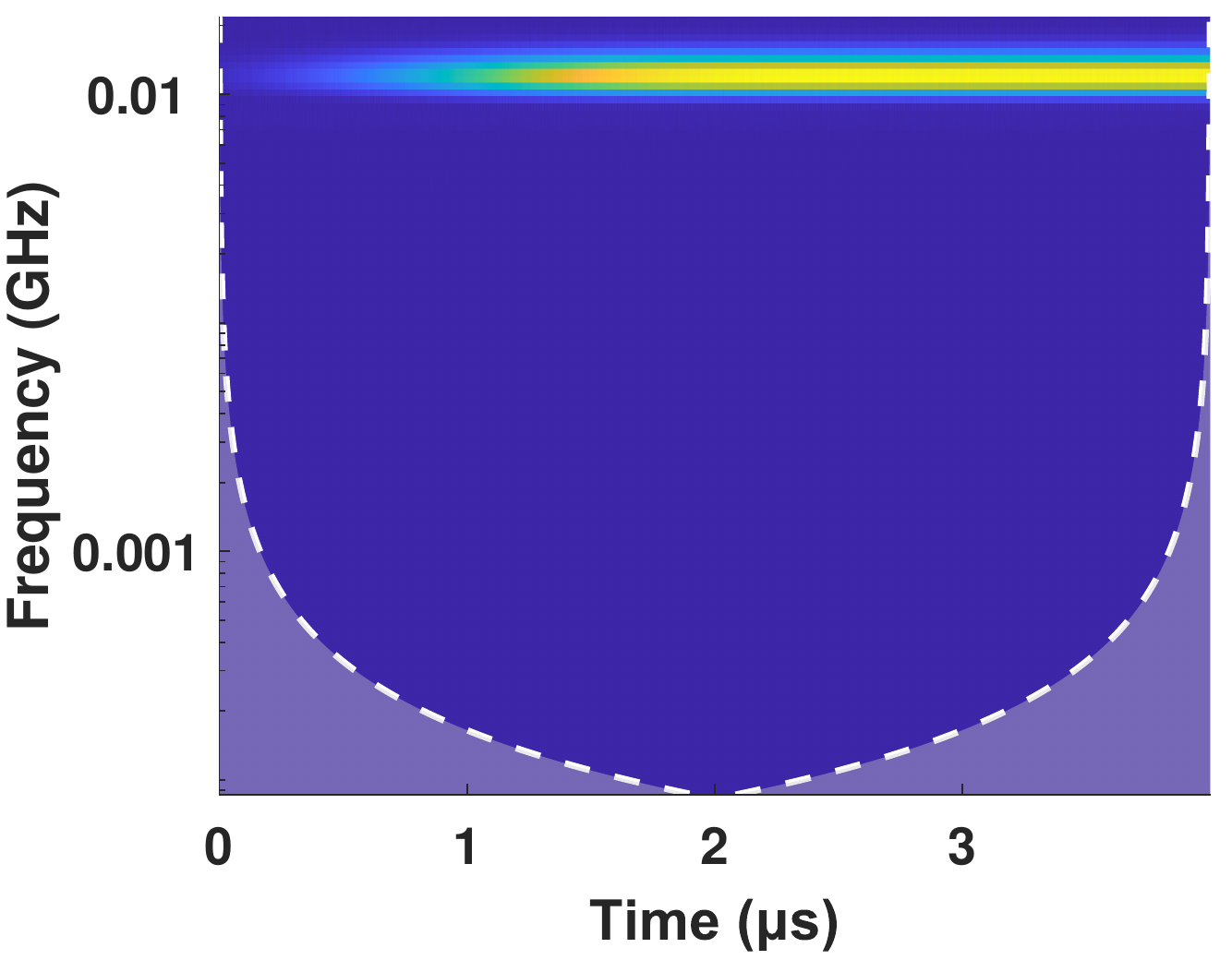}\label{}}
\end{subfloat}
\begin{subfloat}[]{\includegraphics[scale=0.35]{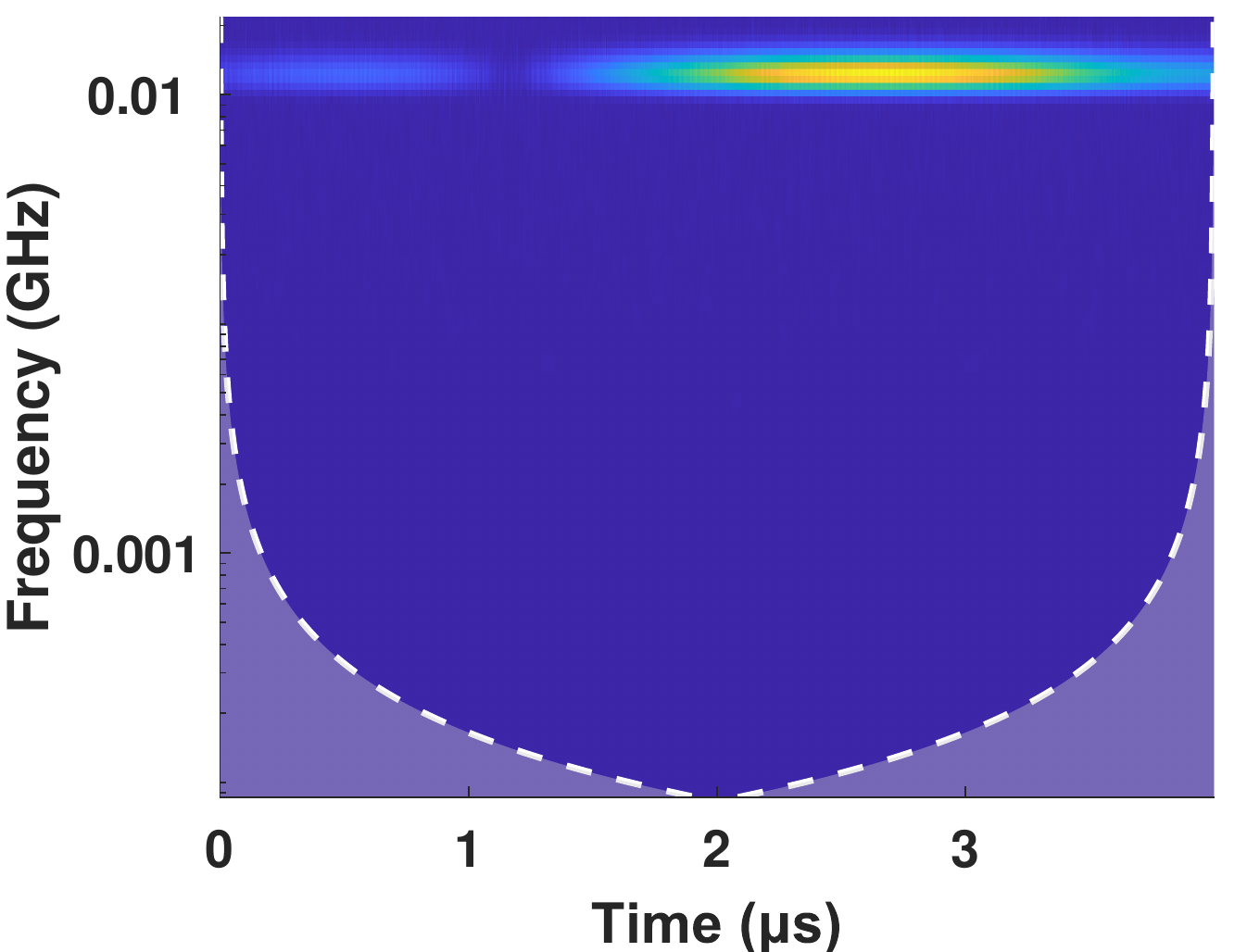}\label{}}
\end{subfloat}
\begin{subfloat}[]{\includegraphics[scale=0.35]{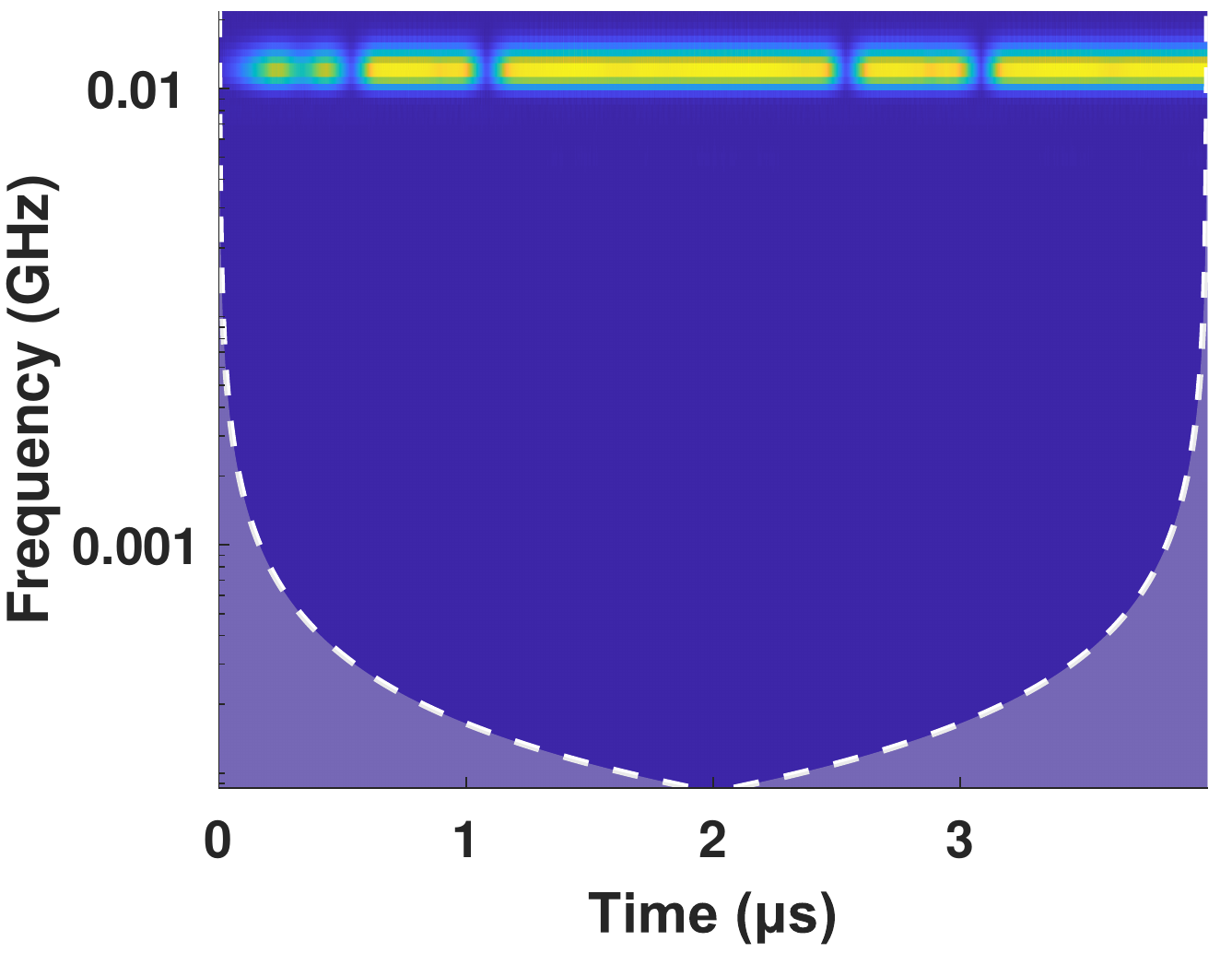}\label{}}
\end{subfloat}
 \caption{{Examples of scalogram extracted from the transient state of the captured RF signals: (a) IPhone 6S, (b) DJI Inspire, and (c) TPLink (WiFi device). }}
  \label{cwt_scalogram_transient}}
  \vspace{-5mm}
\end{figure*}

\begin{figure*}
\center{
\begin{subfloat}[]{\includegraphics[scale=0.35]{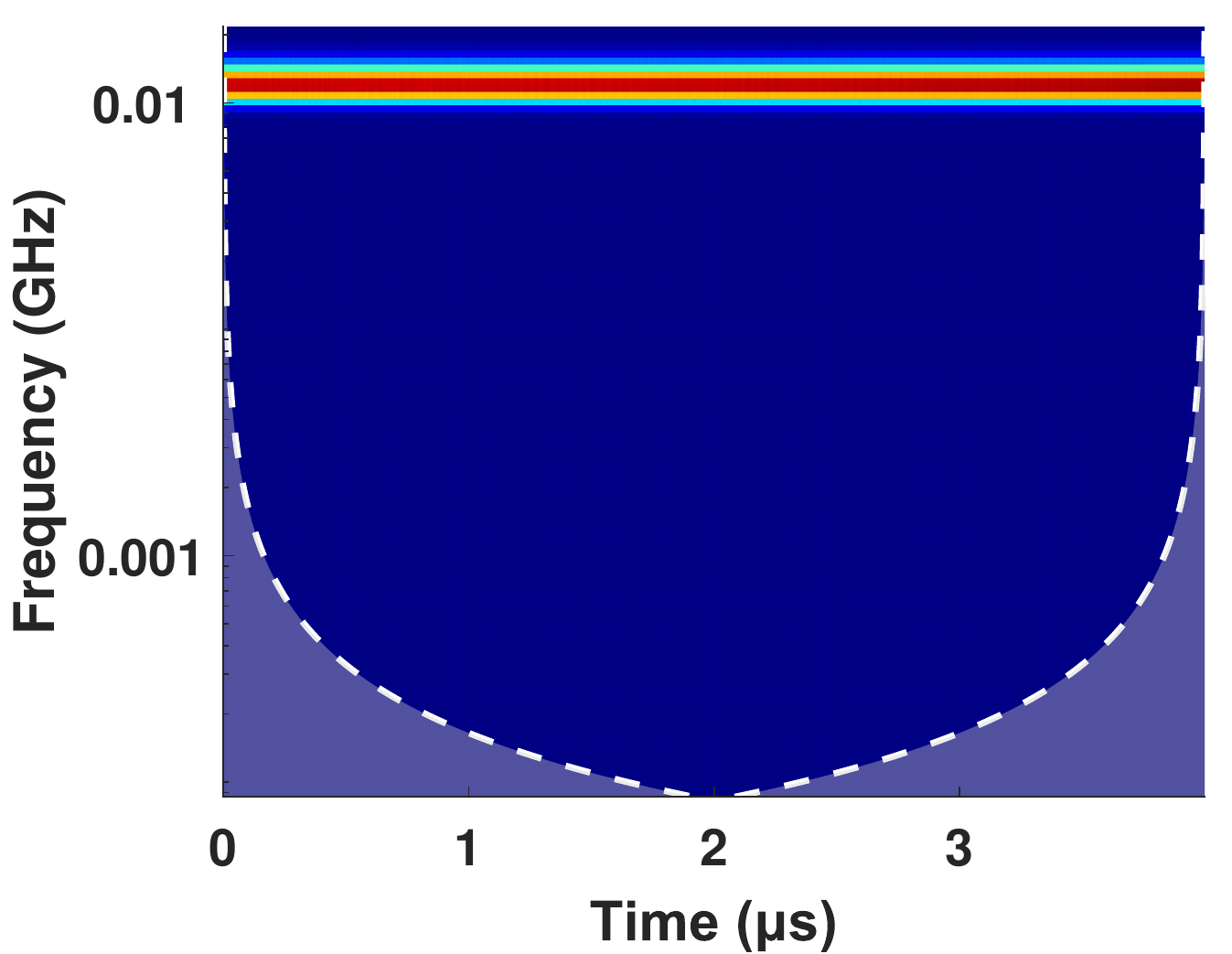}\label{}}
\end{subfloat}
 \begin{subfloat}[]{\includegraphics[scale=0.35]{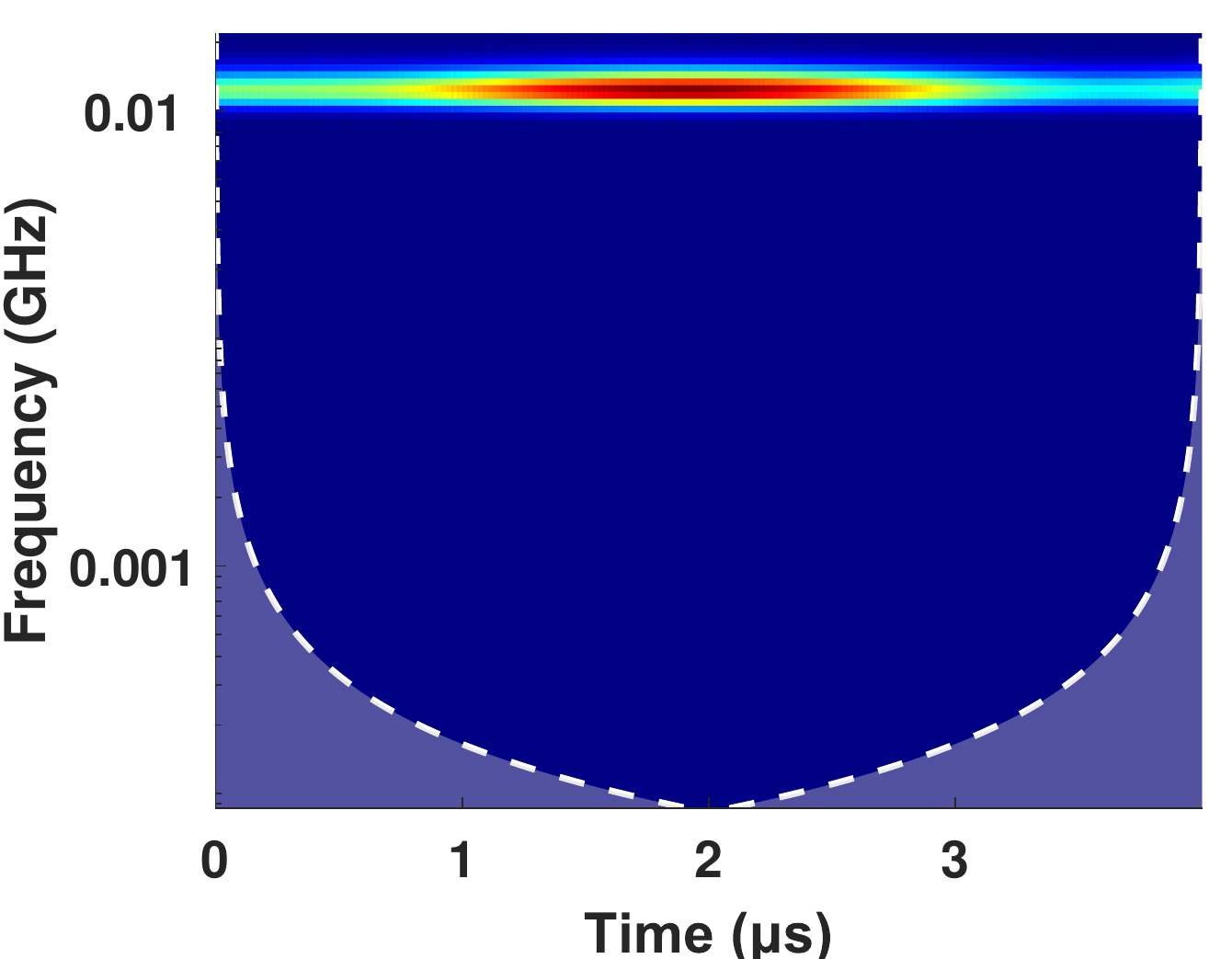}\label{}}
 \end{subfloat}
 \begin{subfloat}[]{\includegraphics[scale=0.35]{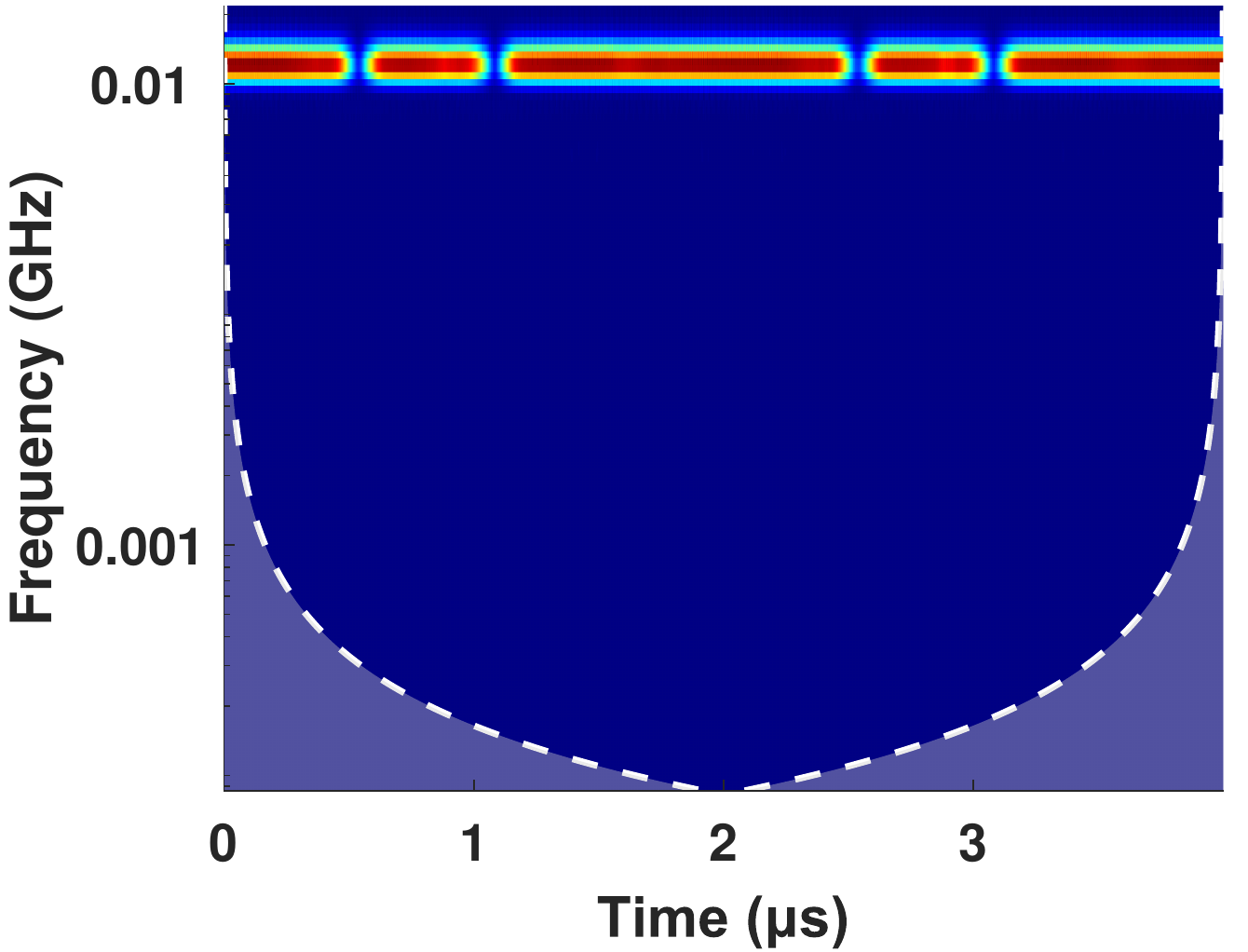}\label{}}
\end{subfloat}

 \caption{{
 Examples of scalogram extracted from the steady state of the captured RF signals: (a) IPhone 6S, (b) DJI Inspire, and (c) TPLink (WiFi device). }}
  \label{cwt_scalogram_steady}}
  \vspace{-5mm}
\end{figure*}

Two categories of features are extracted from the CWT of the signal. These are the transform's coefficients and the scalogram (i.e., an image). We transformed our signals using the implementation of the CWT function in MATLAB. The real part of the transform's coefficients is averaged and transposed as a feature set for the signal representation. As a result of the transformation and averaging, a set of 114 features is acquired as the signature for each signal and this feature set is used to train the ML algorithms. Similarly, we generated the scalogram of the signal as a signature from CWT which is then used to train CNN based algorithm (SqueezeNet). Fig.~\ref{cwt_scalogram_transient} and Fig.~\ref{cwt_scalogram_steady} show the examples of some scalograms generated from a UAV controller (i.e.,DJI Inspire),  WiFi and Bluetooth devices with respect to the transient and steady states of the RF signal.

\subsection{Feature Extraction based on Wavelet Scattering Transform}

WST is an improved time-frequency analytical technique for signals which are based on the wavelet transform. It allows the derivation of low-variance features from signals which can be used by a classifier to discriminate signals \cite{anden2014deep,bruna2013invariant, mallat2012group,mallat2016understanding}.

\begin{figure}
\center{\includegraphics[scale=0.5]{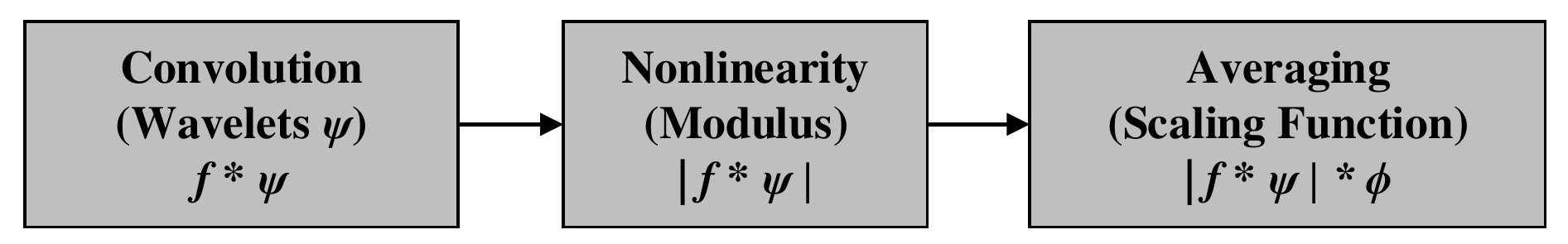}}
\caption{Wavelet scattering transform process flow with the three successive steps in transforming a signal.}
\label{Fig:wst_process}
 \vspace{-5mm}
\end{figure}

The three key factors of WST are \cite{anden2014deep}:
\begin{itemize}
  \item invariance or insensitivity to translation of signals,
  \item stability to signal deformation,
  \item discriminative signals (i.e., informative feature representation).
\end{itemize}
The moving average of the signal enhances invariance and stability. On the other hand, the modulus of a wavelet transform is stable and discriminative. WST exploits these two concepts for an improved time-frequency analysis. The framework of WST is synonymous with the CNN architecture for feature extraction but the computational process does not involve the learning of parameters \cite{anden2014deep}. WST involves three successive processing of signals as shown in Fig.~\ref{Fig:wst_process}. These are the convolution of the wavelet function with a signal, non-linearity by applying modulus operation and low pass filtering averaging using a scaling function.

\begin{figure*}
\center{\includegraphics[ clip,scale=0.83]{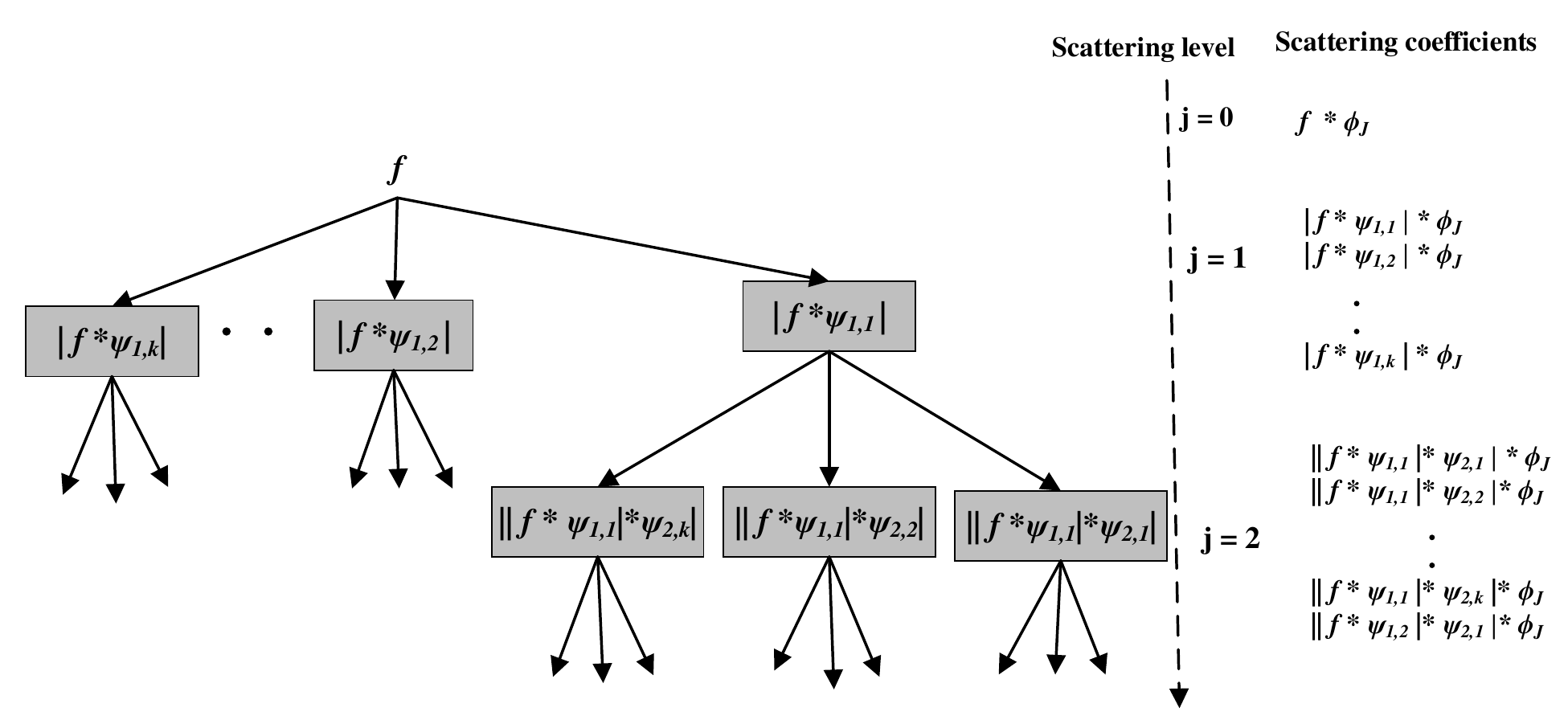}}
\caption{Tree of wavelet scattering transform algorithm for signal decomposition with mathematically expressions for computing the scalogram and scattering coefficients at each scattering levels. }
\label{wst_tree}
 \vspace{-5mm}
\end{figure*}

The wavelet function in wavelet scattering transform is a dilated mother wavelet $\psi$ with a scaling factor ${2}^{-j}$, where $j$ varies from 1 to $J$ (i.e., maximum scattering level) as follows:
\begin {equation} \label{eq:12}
\psi_{j,k}(u) ={2}^{-2j} \psi( {2}^{-j}u ).
\end{equation}
Fig.~\ref{wst_tree} shows the tree algorithm for the WST framework. The node of the tree contains the scalogram coefficients while the scattering coefficients are the convolution of the scalogram coefficient and the scaling function. The first step in WST is to get the translation invariance coefficients by averaging the input signal $f$ with a low pass filter (scaling function) $\phi_J$. These translation invariance coefficients are the zeroth-order scattering coefficients and it is denoted as $S_{0}$:
\begin {equation} \label{eq:13}
S_{0} = f * \phi_J.
\end{equation}
The subsequent steps involve the convolution of the input signal with each wavelet filter $k$ in the first filter bank, followed by the non-linearity operation. This is accomplished by taking the modulus of each of the filtered outputs. This results in the nodes at the first level (i.e., $j=1$) which are the scalogram for the first level, $U_1$, as:
\begin {equation} \label{eq:14}
U_{1} = |f * \psi_{1,k}|.
\end{equation}
The averaging of each modulus with the scaling function will yield the first-order scattering coefficients, $S_1$, as follows:
\begin {equation} \label{eq:15}
S_{1} = |f * \psi_{1,k}| * \phi_J.
\end{equation}
This iterative process is carried out on the next $jth$ level using each wavelet filter $k$ in the  $jth$ filter bank.

\begin{figure*}
\center{
\begin{subfloat}[]{\includegraphics[scale=0.5]{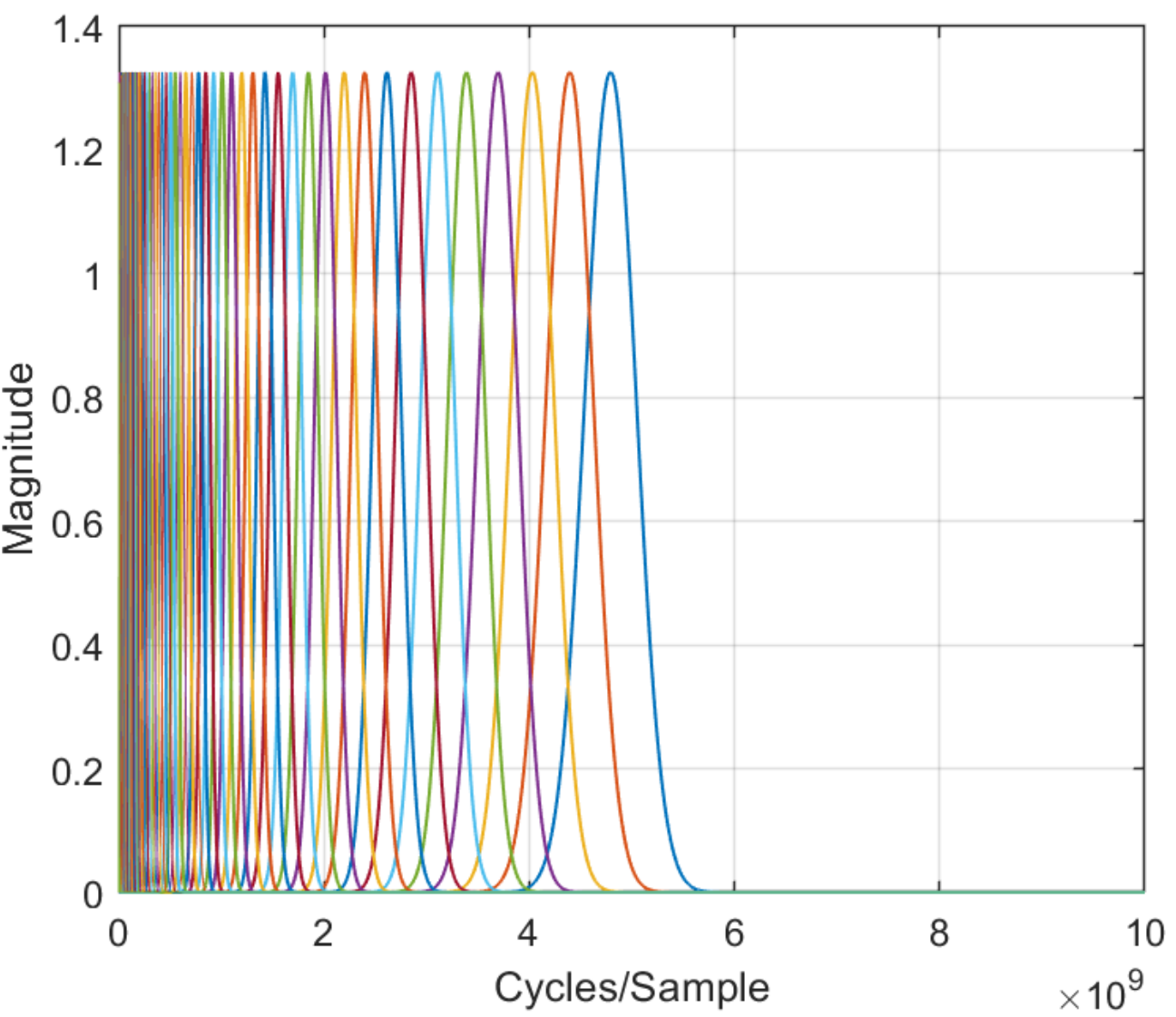}\label{}}
\end{subfloat}
\begin{subfloat}[]{\includegraphics[scale=0.5]{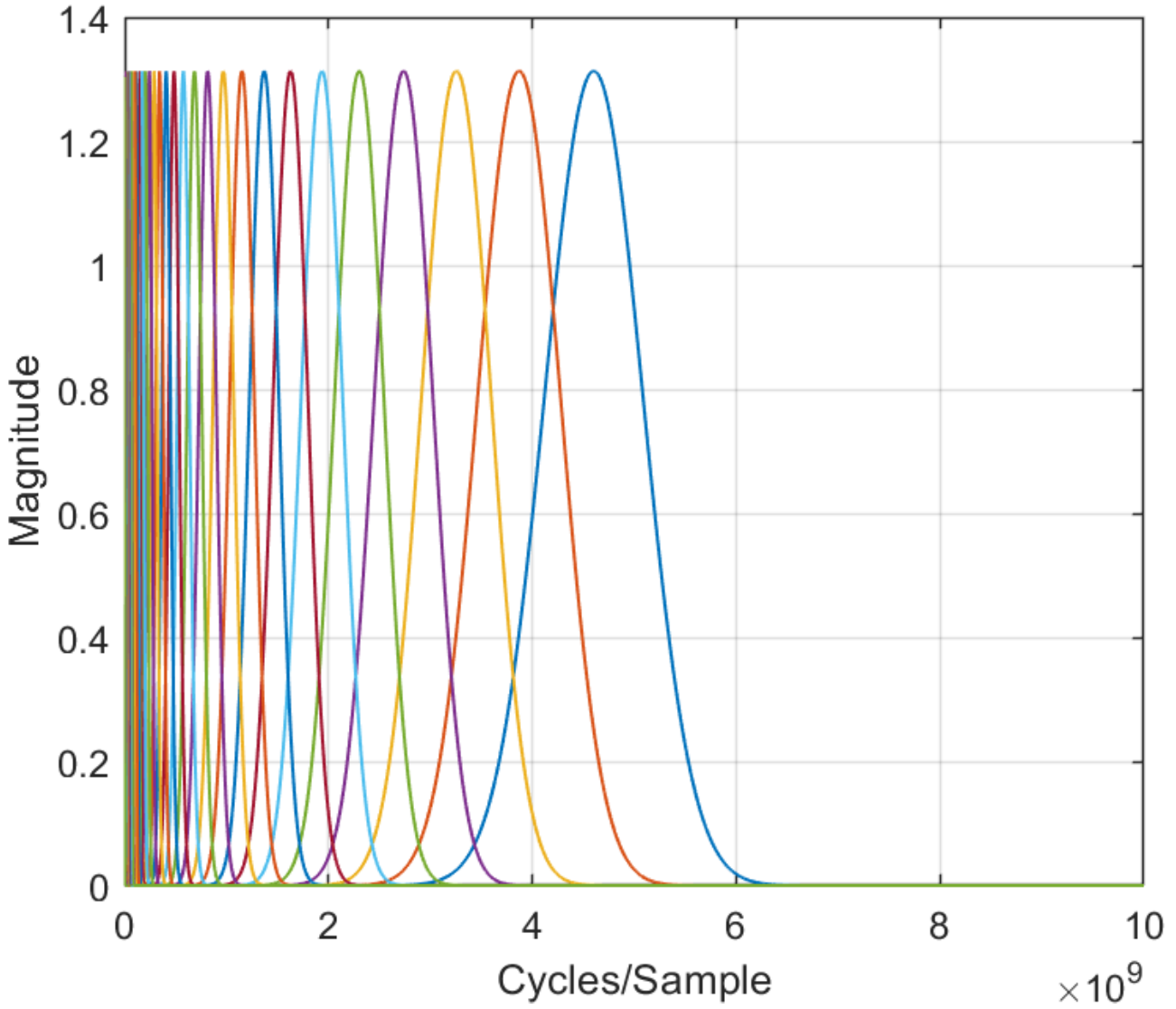}\label{}}
\end{subfloat}
  \caption{{Filter bank for the WST framework: (a) First filter bank, and (b) Second filter bank. }}
  \label{Fig:filterbank}}
  \vspace{-5mm}
\end{figure*}

Similar to CWT, we proposed two feature extraction methods using WST which are the coefficient based feature set and image-based signature (scattergram). The WST framework uses the Gabor (analytic Morlet) wavelet. The Gabor wavelets provide an optimal resolution in both time and frequency domains which optimally extract local features in signals or images \cite{bruna2013invariant,shen2006review}. The implementation of the WST algorithm in MATLAB R2020b is adopted for this work. In WST, as we iterate through the scattering levels, energy is dissipated so two scattering levels are sufficient for systems that use the framework \cite{bruna2013invariant}. This is because experimental results in literature have shown that the third level scattering coefficients can have energy below one percent \cite{bruna2013invariant}. Hence, we propose a two wavelet filter bank for our WST framework.  Fig.~\ref{Fig:filterbank}(a) and Fig.~\ref{Fig:filterbank}(b) show the Gabor wavelets in the first and the second filter bank respectively which are used to compute the scattering coefficients at each level. The quality factors for the first and the second filter bank are eight and four, respectively.

\begin{figure*}{}
\center{
\begin{subfloat}[]{\includegraphics[scale=0.35]{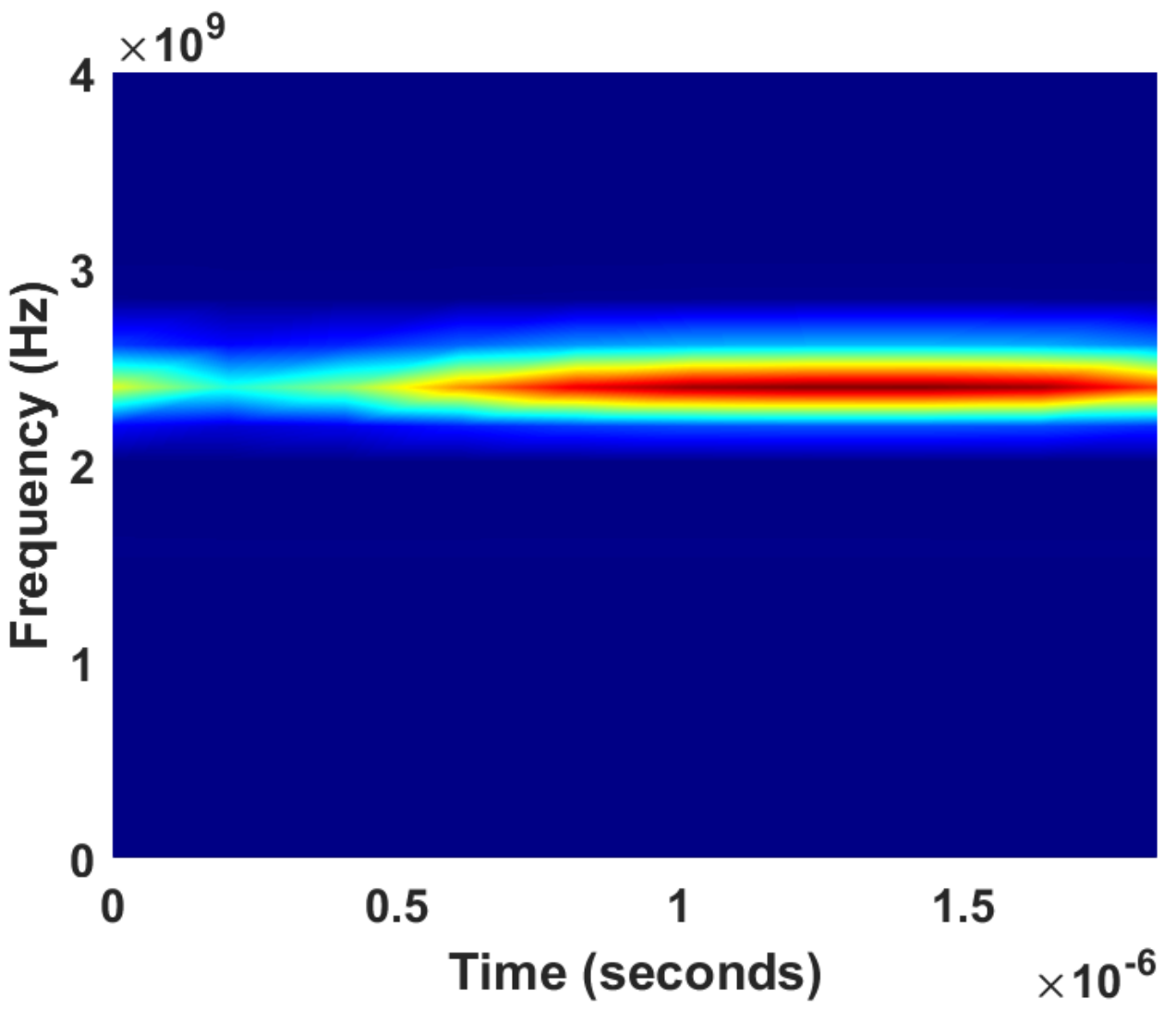}\label{}}
\end{subfloat}
 \begin{subfloat}[]{\includegraphics[scale=0.35]{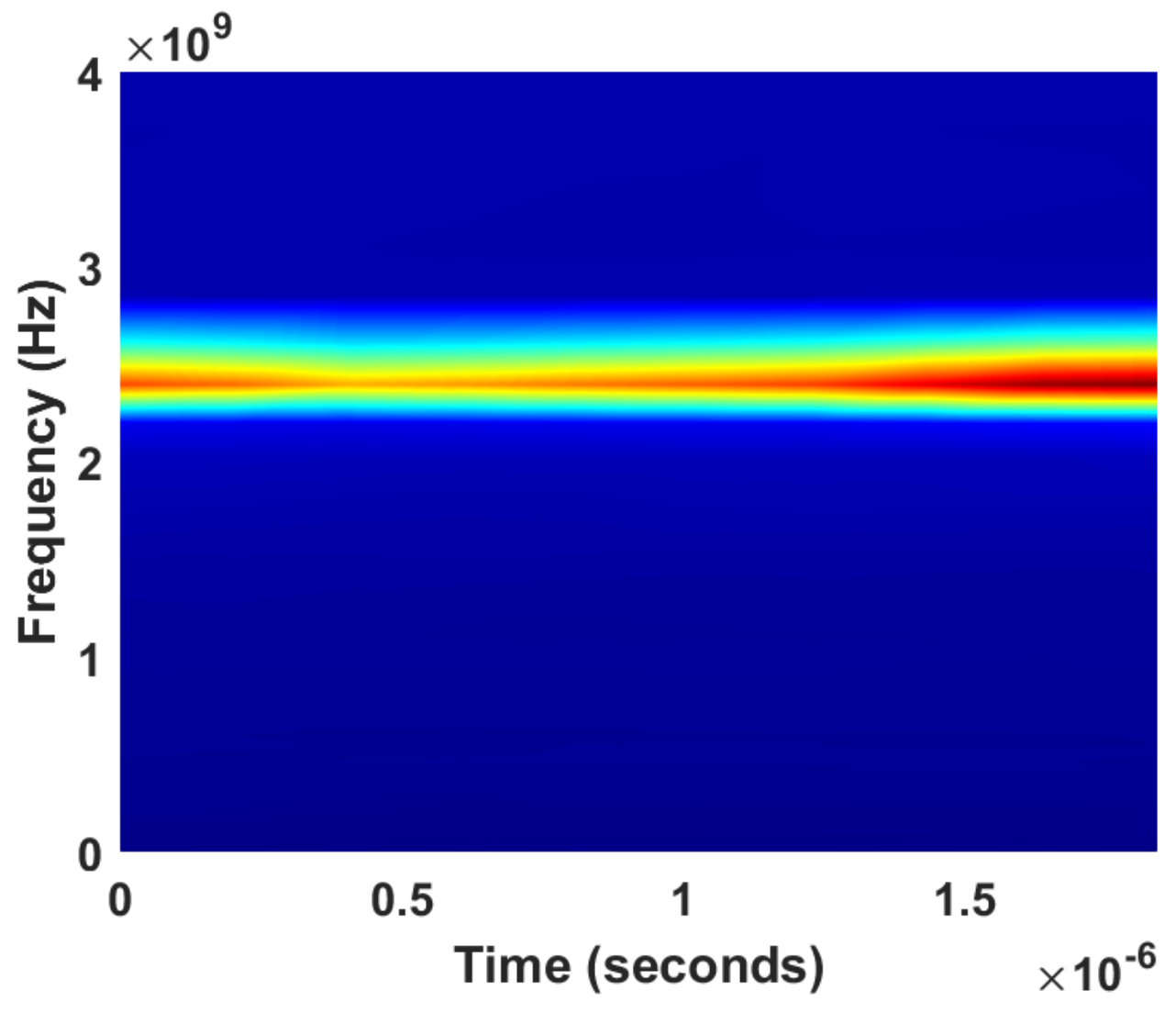}\label{}}
 \end{subfloat}
\begin{subfloat}[]{\includegraphics[scale=0.35]{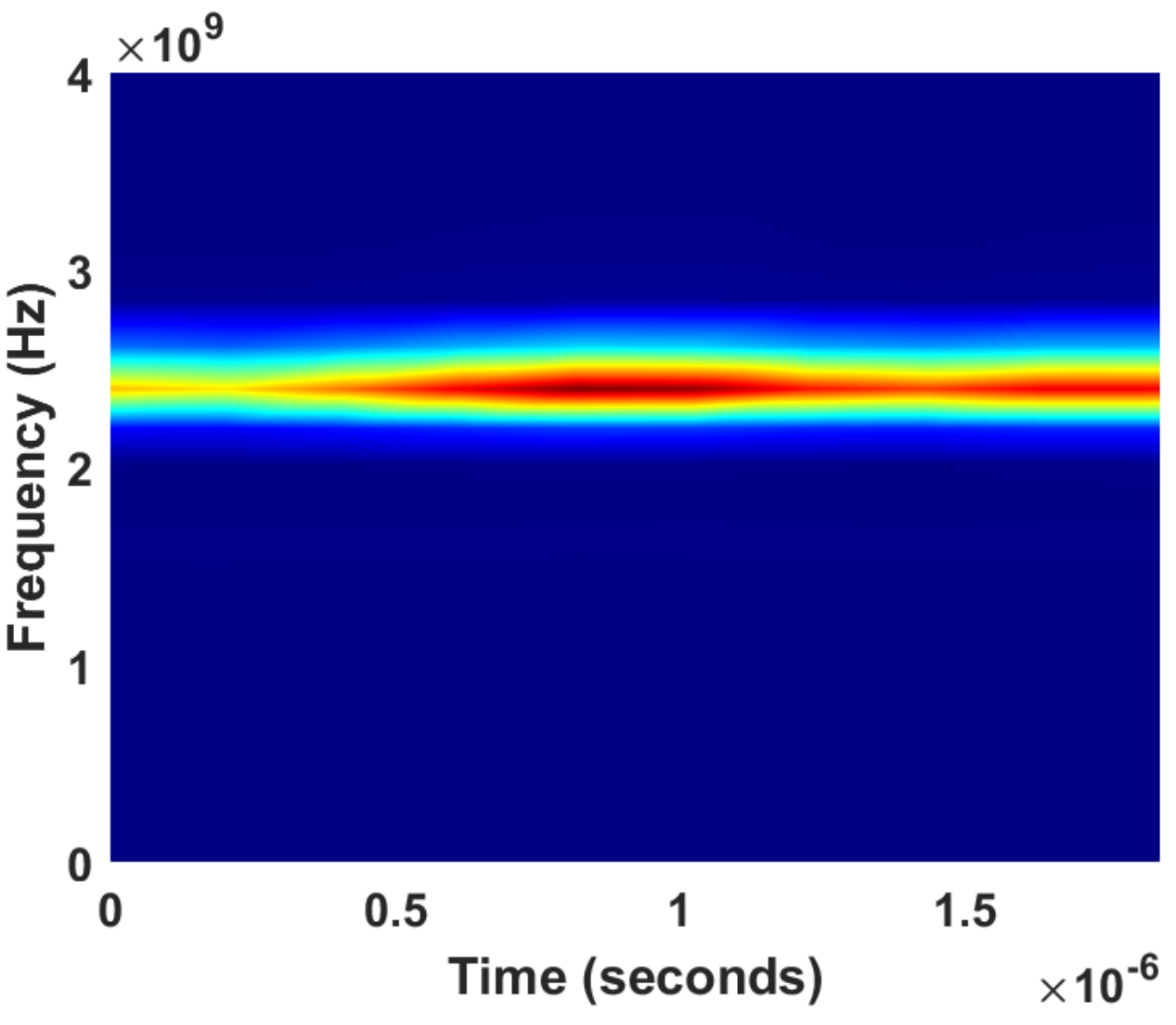}\label{}}
\end{subfloat}


 \caption{{Examples of scattergram extracted from the transient state the captured RF signals: (a) IPhone 6S, (b) DJI Mavic Pro 1, and (c) TPLink (WiFi device).  }}
  \label{wst_scattergram_transient}}
  \vspace{-5mm}
\end{figure*}

\begin{figure*}{}
\center{

\begin{subfloat}[]{\includegraphics[scale=0.35]{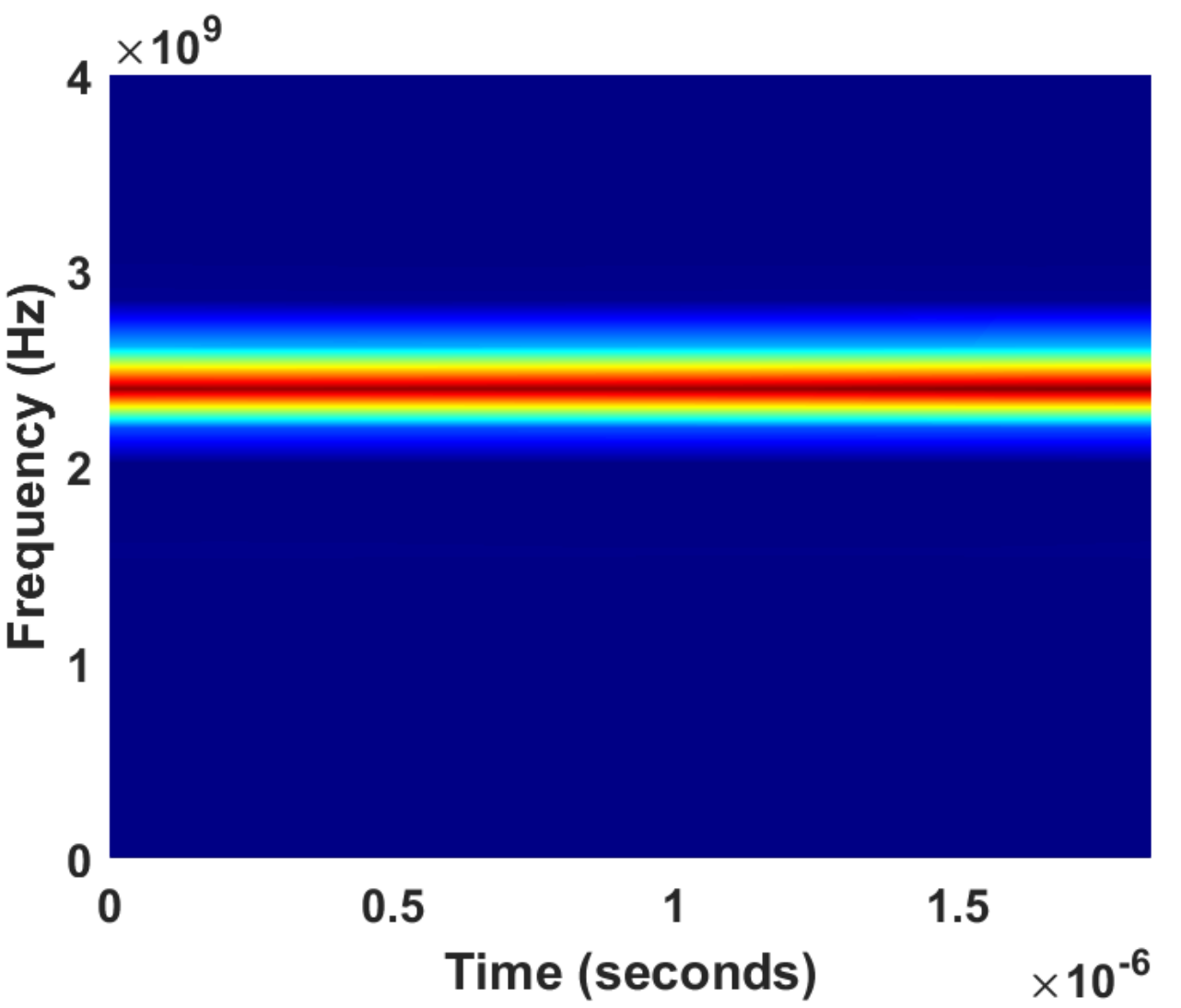}\label{}}
\end{subfloat}
\begin{subfloat}[]{\includegraphics[scale=0.35]{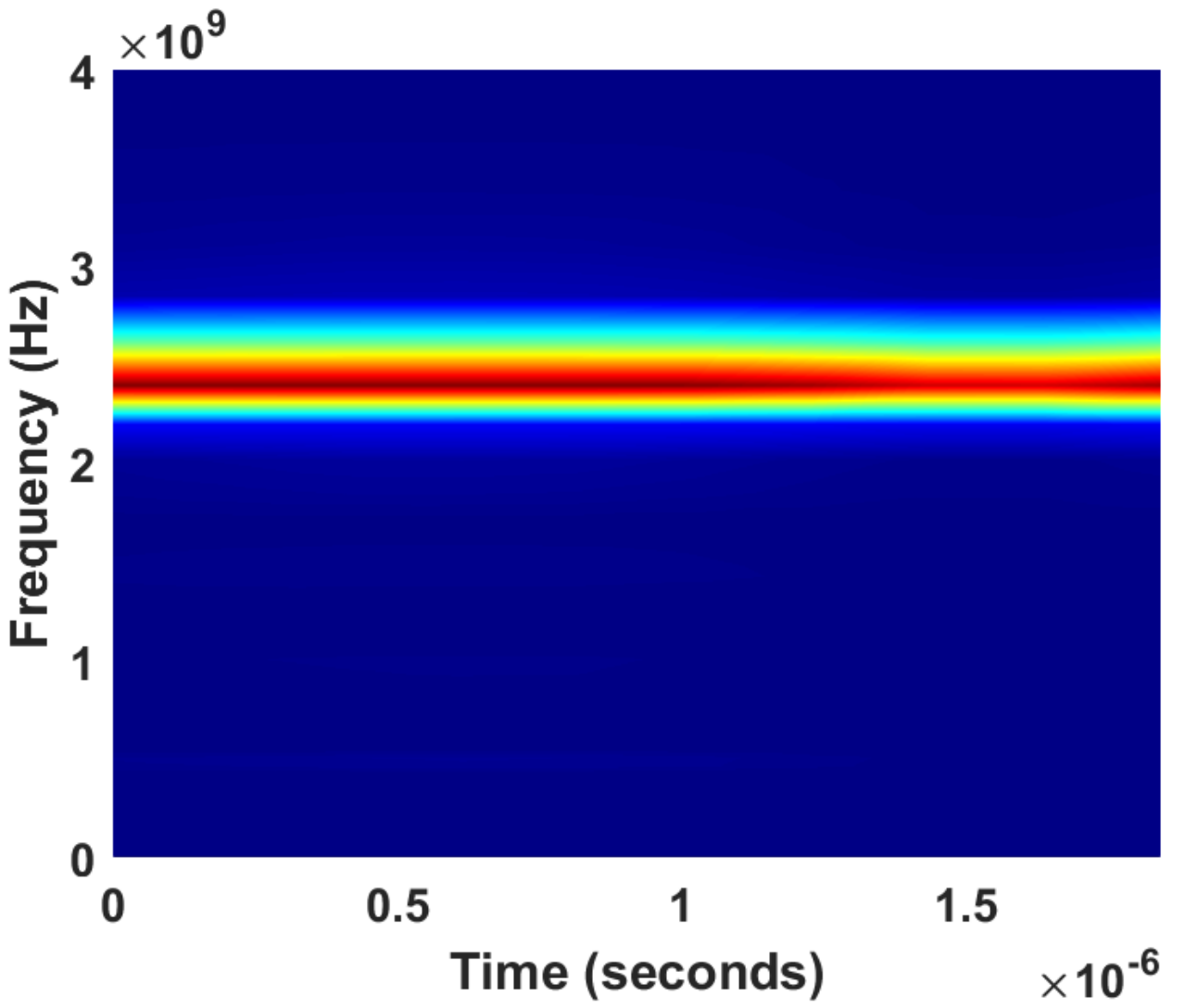}\label{}}
\end{subfloat}
\begin{subfloat}[]{\includegraphics[scale=0.35]{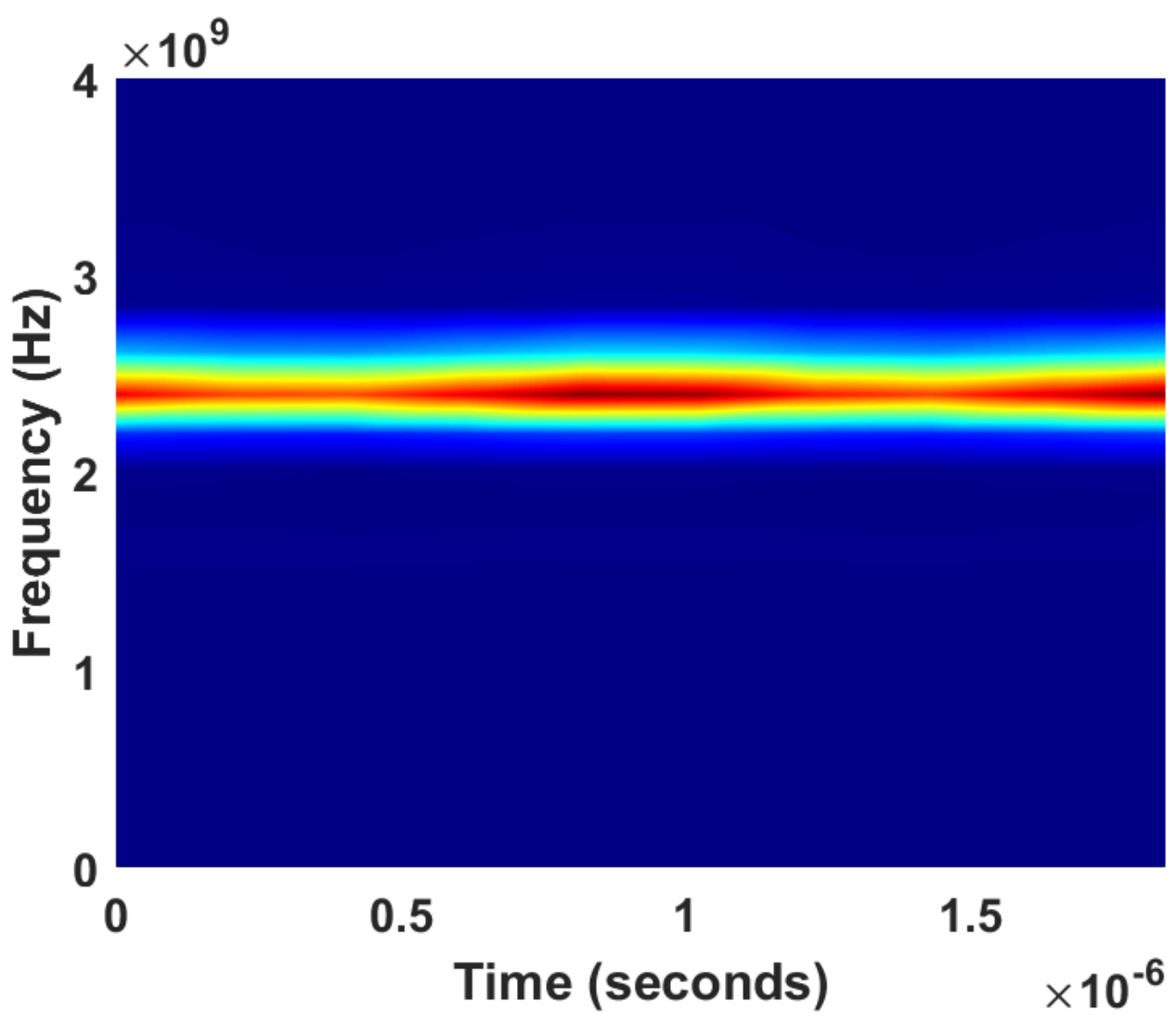}\label{}}
\end{subfloat}

 \caption{{Examples of scattergram extracted from the steady state the captured RF signals: (a) IPhone 6S, (b) DJI Mavic Pro 1, and (c) TPLink (WiFi device). }}
  \label{wst_scattergram_steady}}
  \vspace{-5mm}
\end{figure*}

We average each scattering coefficient generated by the WST framework and this resolved to a total of 1376 features set for each signal. This feature set is used to train the ML algorithm for classification purposes. More so, for the image-based signature, we extracted the scattergram of the scalogram coefficients from the first filter bank to train a pre-trained CNN model called SqueezeNet. Fig.~\ref{wst_scattergram_transient} and Fig.~\ref{wst_scattergram_steady} depict the examples of some scattergrams generated from a UAV controller (i.e.,DJI Mavic Pro 1), WiFi and Bluetooth devices based on state of the RF signal used for feature extraction.

\section{UAV Classification Algorithm}\label{six}

Most of the DDIs system in the literature use statistical and ML techniques to make data-driven detection or classification. ML and DNN (also known as deep learning) are widely used in pattern recognition and provide a means of classifying UAVs based on the patterns in their signatures. ML and DNN algorithms are tools that help in finding a mathematical function or rule-based function that can mimic a given system using n-dimensional input space data, where n is the number of features.

Given $y=F(x)$ from a training set $T={(\vec{x}_1,y_1 ),(\vec{x}_2,y_2 ),(\vec{x}_3,y_3 ),…,(\vec{x}_m,y_m)}$ where $\vec{x_i}$ is feature vectors and  y is the label. If  $y \in C$;  and where $C$ is a finite space, then determining $y$ is a classification problem. Then we can derive the function $F(x)$ as a model from the training set $T$ with $m$ number of samples to map future feature vectors.

The DDI problem can be modeled as a classification problem where the RF signature of a UAV is used for identifying the UAV. Classical ML algorithms such as \textit{k}NN, SVM and ensemble are used for classification in this work. We also use SqueezeNet which is a CNN based algorithm  for classification.

SqueezeNet is a pre-trained model with a small CNN architecture designed to reduce the number of parameters \cite{iandola2016squeezenet}. It has 18 layers and the model can classify images into 1000 classes (i.e., animals, pencil, mouse, etc.) Rather than rebuilding a model from scratch, the concept of transfer learning \cite{pan2009survey} was adopted by using SqueezeNet because the network has learned diverse feature representations from over 1000 images. The architecture of the algorithm was configured for the number of classes in our data set.

\begin{figure}
\center{\includegraphics[scale=0.5]{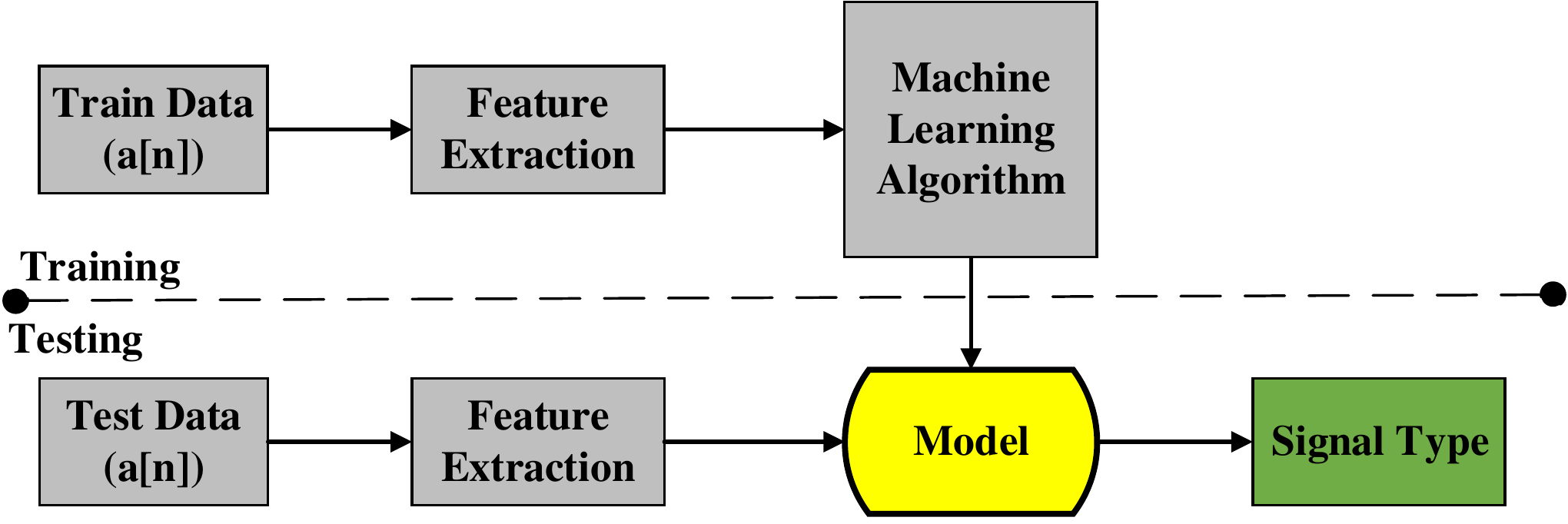}}
\caption{A flowchart of the machine learning model.}
\label{Fig:ml_model_cycle}
\vspace{-5mm}
\end{figure}

The ML modeling aspect of this work follows the flowchart in Fig~\ref{Fig:ml_model_cycle}. The training data (captured RF signal) is first decomposed using a single level HWD as discussed in Section \ref{HWD_section} and the approximation coefficient $a[n]$ is used for extracting the corresponding RF signature or feature sets of the signal. After the feature extraction, the extracted features are used to train an ML algorithm. Similarly, the raw test data is decomposed to compute the approximation coefficients, and using the same feature extraction technique as the training set, the extracted feature set is passed into a trained model for classification of signal type.

Three classical ML algorithms are used to train coefficient based signatures in this work. These are SVM, \textit{k}NN and ensemble. Because the dimensionality of the coefficient based features is quite large (i.e., CWT has 114 features and WST has 1376 features), we apply principal component analysis (PCA) to reduce the dimensionality. On the other hand, SqueezeNet was used to train image-based signatures.

The training of the models is done in MATLAB R2020b using the classification learner and deep network designer frameworks for both ML algorithms and  SqueezeNet respectively. We experimented on the university of Louisville Cybersecurity Education data center where a Common KVM  4 Core system with 300 GB RAM and 120 GB storage was used.

Rather than using the grid search technique which is computationally tedious for hyperparameter tuning or randomized search which is a random process, Bayesian optimization is adopted for the hyperparameter tuning of the ML algorithms to obtain the best hyperparameters for the classifiers. This optimization method is based on the "Bayes' theorem" which takes advantage of the prior observations of the classifier's loss, to determine the next optimal hyperparameters that will reduce the loss \cite{snoek2012practical}.

\section{Experimental Results and Discussions} \label{seven}
We present the results of each trained model by using the test data to evaluate the performance of the model. We focus our results on seven main directions by comparing and contrasting the results of the four proposed feature extraction approaches. These directions are:
\begin{itemize}
\item the state of the RF signal used for feature extraction;
\item the behavior of the model when the classification is a three-class problem (i.e., classifying the signals to Bluetooth, WiFi, and UAV signal) and when the model is a ten class problem (i.e., identifying each device);
\item  the use of coefficient based signature over an image-based signature;
\item the impact of reducing the dimensionality of coefficient based signature on model performance;
\item model performance under varying SNR;
\item computational time complexity of feature extraction methods and the classifier's inference time;
\item Lastly, the performance of the type of waveform transforms used for feature extraction.
\end{itemize}

In a classification problem, there are several metrics for evaluating the performance of a classifier. These include accuracy, precision, recall, and $F_1$-score. Using accuracy can misrepresent the performance of a classifier especially when the test data is skewed, so a confusion matrix is used in addition. Because of space limitation, we presented some of the confusion matrices of our models, and also, where the confusion matrix is not shown, we provided the precision, recall, and $F_1$-score of the model, which give the summary of the confusion matrix of the model for a thorough understanding of the model performance. The definition of these metrics is given below

\begin {equation} \label{eq:7}
\text{Accuracy}=\frac{ T_{\rm{P}}+T_{\rm{N}}} {T_{\rm{P}}+T_{\rm{N}}+F_{\rm{P}}+F_{\rm{N}}},
\end{equation}

\begin {equation} \label{eq:8}
\text{Precision}= \frac{T_{\rm{P}}}{T_{\rm{P}}+F_{\rm{P}}},
\end{equation}

\begin {equation} \label{eq:9}
\text{Recall}= \frac{T_P}{T_P+F_N},
\end{equation}

\begin{equation} \label{eq:10}
F_{\text{1}}~\text{score}= 2\left(\frac{Precision \times Recall}{Precision + Recall}\right),
\end{equation}
where $T_\text{P}$, $T_\text{N}$, $F_\text{P}$ and $F_\text{N}$ represent true positive, true negative, false positive and false negative, respectively.

\subsection{Model Performance based on Classification Metrics}

Table~\ref{Table_classificationAccuracy} shows the average accuracy of the classifiers using the proposed feature extraction methods under the two signal states (i.e, transient and steady states) at 30~dB SNR. While we focus on the classification accuracy as the main metric, we provided recall, precision, and $F_1$-score values for the classifiers in Table~\ref{Table_classificationrecall}, Table~\ref{Table_classificationprecision}, and  Table~\ref{Table_classification_f1score}, respectively, as additional information to show that the classifiers generalize well and that there are no forms of bias due to unbalanced data (i.e., in group-device classification) in our classification accuracy. For any given classifier, the four performance metrics are within the same range.

From Table~\ref{Table_classificationAccuracy}, using the coefficients of CWT of the signal irrespective of the signal states, in determining the three types of device (UAV controller, Bluetooth, and WiFi), the ML algorithms and SqueezeNet yield an average classification accuracy of over $99\%$. \textit{k}NN outperforms other ML algorithms (i.e, SVM and ensemble) and SqueezeNet with an average classification accuracy of $99.8\%$. This shows that either of the states has sufficient information in classifying the device type. Because the row-wise averaging of the coefficients of CWT for each signal resulted in a 114 feature set and overloading of data through high dimensionality could restrain the performance of a model \cite{kantardzic2011data}. Hence, in an attempt to remove the redundant feature to avoid model overfitting we use PCA for dimensionality reduction of the feature set using $95\%$ explained variance as a constraint. This reduced the feature set from 114 to 1. Using one principal component as a feature set by exploiting the transient state of the signal, the accuracy of the ML models falls to $92.5\%$, $92.9\%$ and $92.9\%$ for \textit{k}NN, SVM, and ensemble, respectively. Likewise, by exploiting the steady state of the signal, a significant drop in accuracy was seen when using the PCA on the CWT coefficients. The accuracy of \textit{k}NN, SVM, and ensemble are  $79\%$, $79.3\%$, and $79.1\%$, respectively. \textit{k}NN outperforms other ML algorithms before applying PCA but it ended with the lowest average accuracy after the dimensionality reduction when exploiting either the transient or steady state. Squeezenet which uses the image-based signature (scalogram) gives an accuracy of $99\%$ and $99.4\%$ for the transient and steady state respectively.

When CWT is used for the identification of each device (i.e., the 10 classes problem) the accuracies of the models are reduced and \textit{k}NN outperforms other ML algorithms using the steady state of the signals with an accuracy of $87\%$. The performance of \textit{k}NN decreased the most when the feature space (i.e., coefficients of CWT) is reduced using PCA to about $57.1\%$. The performance of the ML models significantly declines whenever PCA is applied to the coefficients of CWT irrespective of the signal state or number of classes. On the other hand, the  SqueezeNet gives average accuracy of $88.4\%$ and $77.4\%$ for the transient and steady state respectively.

\begin{table*}
\setlength{\tabcolsep}{1.3pt}
\centering
\caption{Average classification accuracy (\%) of the ML models and SqueezeNet based on the feature extraction methods and the state of RF signal use for fingerprinting at 30 dB SNR.}
\label{Table_classificationAccuracy}
\begin{tabular}{|c|c|c|c|c|c|c|c|c|c|}

\hline
 s/n & Algorithm & \multicolumn{4}{c|}{Continuous Wavelet Transform}  &  \multicolumn{4}{c|}{Wavelet Scattering Transform} \\    \cline{3-10}

 & & \multicolumn{2}{c|} {3 Classes} &  \multicolumn{2}{c|} {10 Classes} &  \multicolumn{2}{c|} {3 Classes} &  \multicolumn{2}{c|} {10 Classes} \\  \cline{3-10}

& & Transient & Steady & Transient & Steady & Transient & Steady & Transient & Steady\\

 \hline
1 & KNN &99.8 & 99.7 & 83.8  & 87.0 & 99.8 & 	99.9 & 	87.9 & 	87.3 \\

\hline
2 & SVM & 99.5 &	99.1 &	81.8 &	80.8 &	99.6 &	99.9 &	88.2 &	87.4 \\

\hline
3 & Ensemble & 99.5 &	99.4 &	85.2 &	84.9 &	99.9 &	99.9 &	90.9 &	85.0 	 \\

\hline
4 & KNN + PCA & 92.5 &	79.0 &	68.8 &	57.1 &	99.9 &	99.8 &	88.3 &	87.2 	 \\

\hline
5 & SVM + PCA & 92.9 &	79.3 &	69.6 &	57.6 &	99.7 &	99.7 &	89.3 &	89.4	 \\

\hline
6 & Ensemble + PCA & 92.9 &	79.1 &	70.0 &	60.7 &	98.9 &	99.7 &	87.4 &	85.9	 \\

\hline
7 &  SqueezeNet & 99.0	& 99.4 & 	88.4 &	77.4 &	99.2 &	99.5 &	88.5 &	76.1 	 \\
\hline
\end{tabular}
 \vspace{-3mm}
\end{table*}

\begin{table*}
\setlength{\tabcolsep}{1.3pt}
\centering

\caption{Average classification recall (\%) of the ML models and SqueezeNet based on the feature extraction methods and the state of RF signal use for fingerprinting at 30 dB SNR.}
\label{Table_classificationrecall}
\begin{tabular}{|c|c|c|c|c|c|c|c|c|c|}

\hline
 s/n & Algorithm & \multicolumn{4}{c|}{Continuous Wavelet Transform}  &  \multicolumn{4}{c|}{Wavelet Scattering Transform} \\   \cline{3-10}

 & & \multicolumn{2}{c|} {3 Classes} &  \multicolumn{2}{c|} {10 Classes} &  \multicolumn{2}{c|} {3 Classes} &  \multicolumn{2}{c|} {10 Classes} \\    \cline{3-10}

& & Transient & Steady & Transient & Steady & Transient & Steady & Transient & Steady\\

 \hline
1 & KNN & 99.7 &	99.5 &	84.3 &	87.3  &	99.8  &	99.8  &	88.1  &	87.7 \\

\hline
2 & SVM & 99.3	& 98.8 &	84.2 &	81.6  &	99.4  &	99.8  &	88.6  &	88.7 \\

\hline
3 & Ensemble & 99.3	& 99.4 &	86.8  &	86.3  &	99.9  &	99.8  &	91.5  &	86.5 	 \\

\hline
4 & KNN + PCA & 89.5 &	72.5  &	71.6  &	55.9  &	99.9  &	99.8  &	88.9 &	88.0 	 \\

\hline
5 & SVM + PCA & 90.1  &	73.0 &	72.1 &	57.1 &	99.5  &	99.6  &	89.8  &	89.9	 \\

\hline
6 & Ensemble + PCA & 89.9 &	72.6 &	73.1 &	60.3  &	98.9  &	99.7  &	88.1  &	86.8	 \\

\hline
7 &  SqueezeNet & 98.5	& 99.0  &	88.8  &	76.5  &	99.0  &	99.2  &	88.7  &	77.2 	 \\
\hline
\end{tabular}
 \vspace{-3mm}
\end{table*}

\begin{table*}
\setlength{\tabcolsep}{1.3pt}
\centering

\caption{Average classification precision (\%) of the ML models and SqueezeNet based on the feature extraction methods and the state of RF signal use for fingerprinting at 30 dB SNR.}
\label{Table_classificationprecision}
\begin{tabular}{|c|c|c|c|c|c|c|c|c|c|}

\hline
 s/n & Algorithm & \multicolumn{4}{c|}{Continuous Wavelet Transform}  &  \multicolumn{4}{c|}{Wavelet Scattering Transform} \\   \cline{3-10}

 & & \multicolumn{2}{c|} {3 Classes} &  \multicolumn{2}{c|} {10 Classes} &  \multicolumn{2}{c|} {3 Classes} &  \multicolumn{2}{c|} {10 Classes} \\   \cline{3-10}

& & Transient & Steady & Transient & Steady & Transient & Steady & Transient & Steady\\

 \hline
1 & KNN & 99.9 &	99.8  &	83.8  &	87	 & 99.8 &	99.9  &	87.9 &	87.3 \\

\hline
2 & SVM & 99.5 &	99.3 &	81.8  &	80.8  &	99.6  &	99.9  &	88.2  &	87.4 \\

\hline
3 & Ensemble & 99.4  &	99.1  &	85.2 &	84.9  &	99.8  &	99.9  &	90.9  &	85.0 	 \\

\hline
4 & KNN + PCA & 90.9  &	74.1  &	68.8  &	57.1  &	99.8  &	99.8  &	88.3  &	87.2 	 \\

\hline
5 & SVM + PCA & 90.8  &	76.8  &	69.6  &	57.6  &	99.7 &	99.7  &	88.3  &	89.4	 \\

\hline
6 & Ensemble + PCA & 91.5 &	75.4  &	70.0 &	60.7  &	98.6  &	99.6  &	87.4  &	85.9	 \\

\hline
7 &  SqueezeNet & 99.3	& 99.7	 & 88.4	& 77.4 &	99.0 &	99.7 &	88.5 &	76.1	 \\
\hline
\end{tabular}
 \vspace{-3mm}
\end{table*}

\begin{table*}
\setlength{\tabcolsep}{1.3pt}
\centering

\caption{Average $F_1$-score (\%) of the ML models and SqueezeNet based on the feature extraction methods and the state of RF signal use for fingerprinting at 30 dB SNR.}
\label{Table_classification_f1score}
\begin{tabular}{|c|c|c|c|c|c|c|c|c|c|}

\hline
 s/n & Algorithm & \multicolumn{4}{c|}{Continuous Wavelet Transform}  &  \multicolumn{4}{c|}{Wavelet Scattering Transform} \\   \cline{3-10}

 & & \multicolumn{2}{c|} {3 Classes} &  \multicolumn{2}{c|} {10 Classes} &  \multicolumn{2}{c|} {3 Classes} &  \multicolumn{2}{c|} {10 Classes} \\    \cline{3-10}

& & Transient & Steady & Transient & Steady & Transient & Steady & Transient & Steady\\

 \hline
1 & KNN & 99.8 &	99.7 &	83.6 &	86.9 &	99.8 &	99.9 &	88.0 &	87.4 \\

\hline
2 & SVM & 99.4 &	99.0 &	82.5 &	81.1 &	99.5 &	99.9 &	88.4 &	87.8 \\

\hline
3 & Ensemble & 99.3	& 99.3 &	85.5 &	85.3 &	99.9  &	99.9  &	91.0  &	85.3 	 \\

\hline
4 & KNN + PCA & 90.1  &	72.7  &	68.3  &	55.3  &	99.9  & 	99.8  &	88.5  &	87.3 	 \\

\hline
5 & SVM + PCA & 90.5  &	74.4  &	69.7 &	55.7  &	99.6 &	99.7  &	89.5 &	89.6	 \\

\hline
6 & Ensemble + PCA & 90.5 &	73.6  &	69.8  &	59.5  & 	98.8 &	99.7 &	87.7 &	86.2	 \\

\hline
7 &  SqueezeNet & 98.9	 & 99.3 &	88.5  &	76.3  &	99.0 &	99.4  &	88.6  &	76.0	 \\
\hline
\end{tabular}
 \vspace{-3mm}
\end{table*}

\begin{figure*}{}
\center{
\begin{subfloat}[]{\includegraphics[scale=0.45]{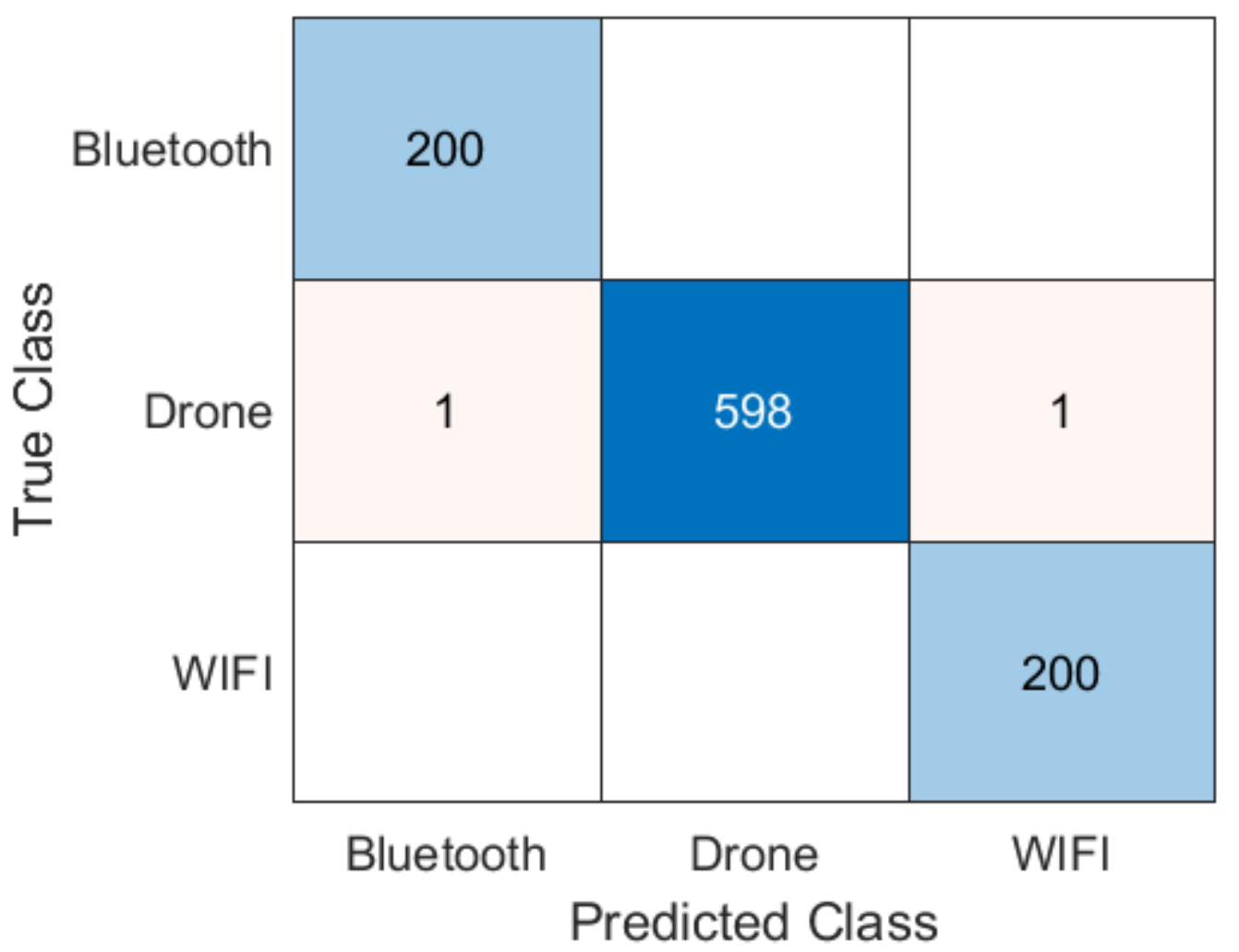}\label{}}
\end{subfloat}
\begin{subfloat}[]{\includegraphics[scale=0.6]{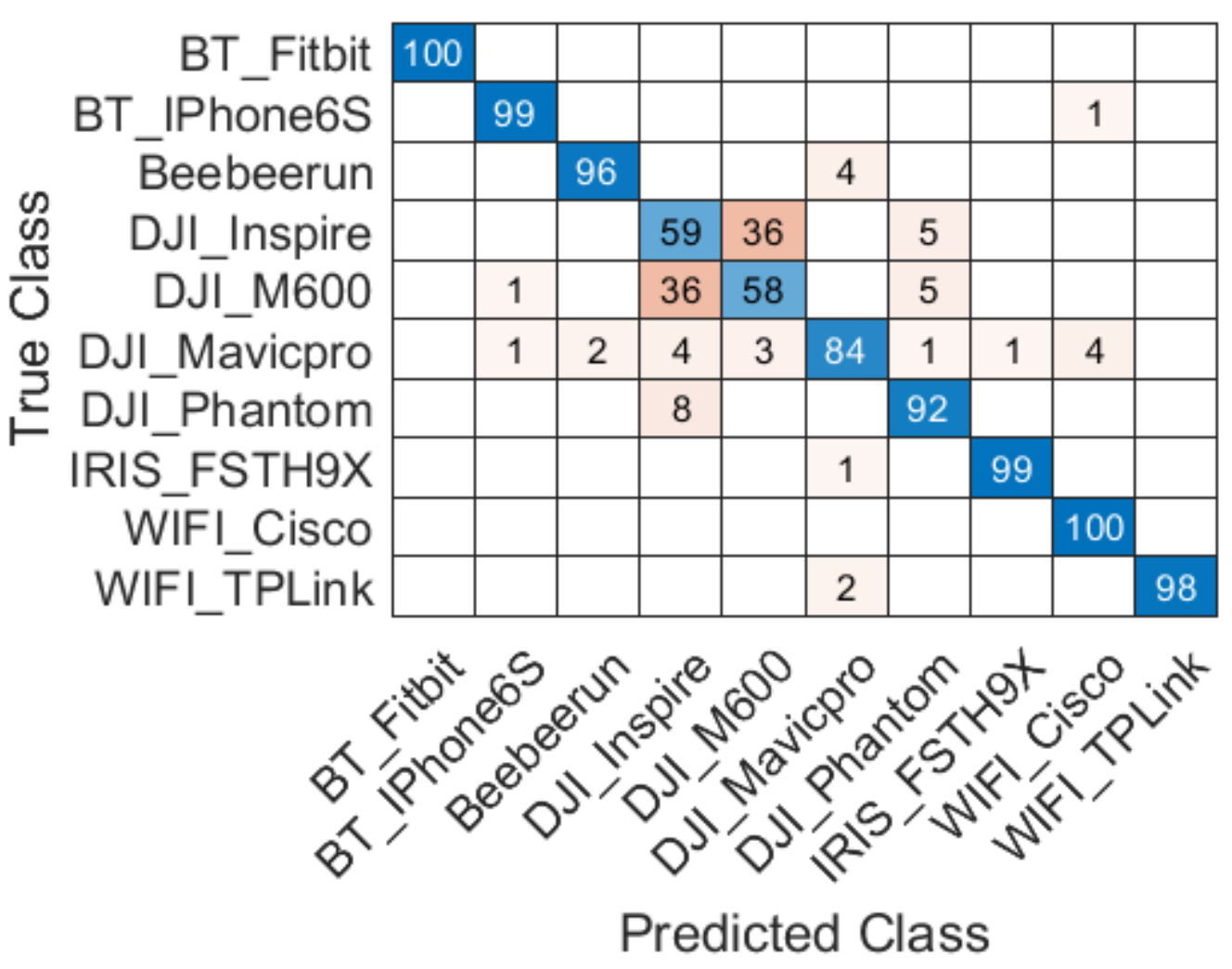}\label{}}
\end{subfloat}

\begin{subfloat}[]{\includegraphics[scale=0.45]{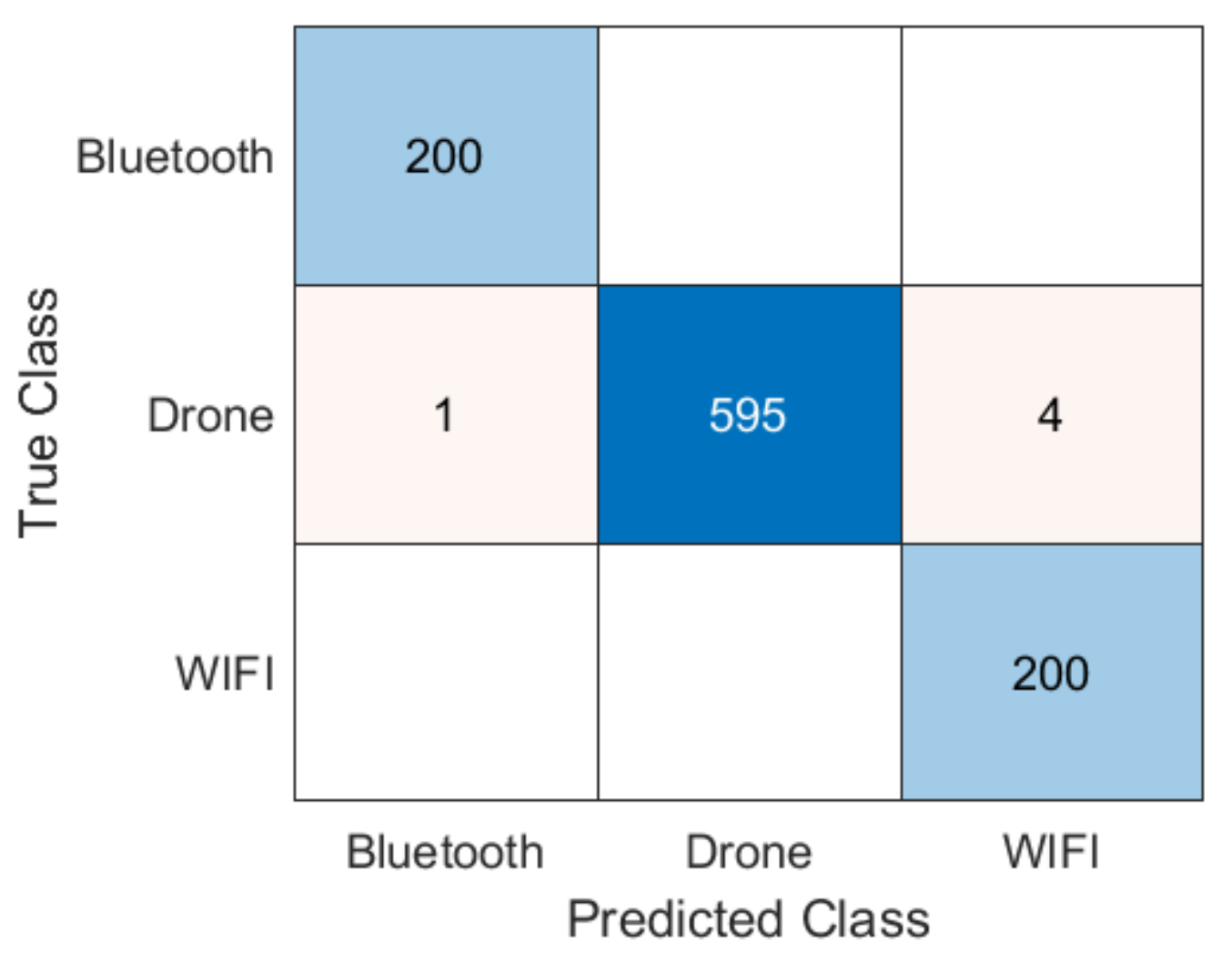}\label{}}
\end{subfloat}
\begin{subfloat}[]{\includegraphics[scale=0.6]{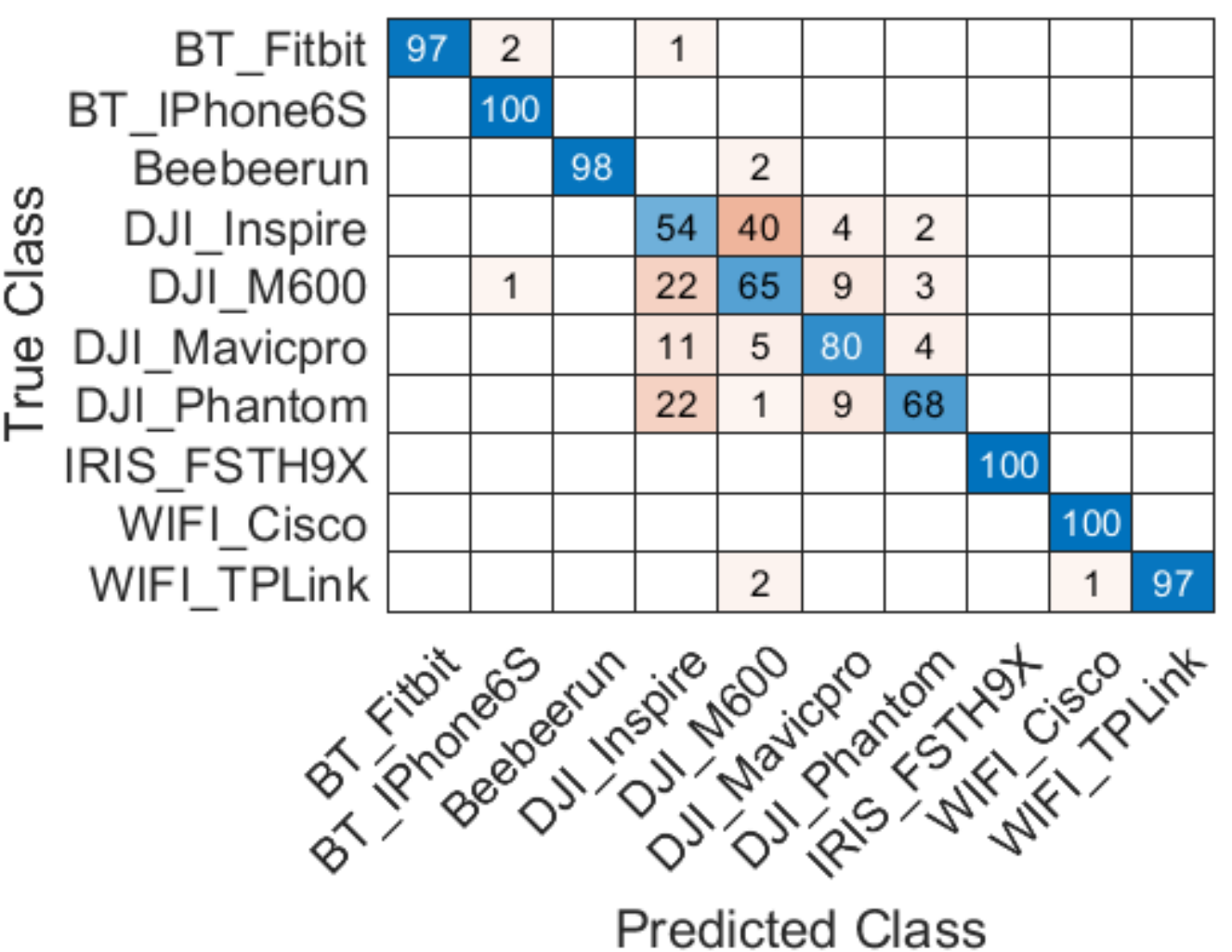}\label{}}
\end{subfloat}

 \caption{{Confusion matrices: (a) Group-device classification using  the transient state, CWT framework and \textit{k}NN,  (b) Specific-device classification using transient state, CWT framework and SqueezeNet (c) Group-device classification using  the steady state, WST framework and SqueezeNet,  and (d) Specific-device classification using transient state, WST framework and Ensemble + PCA.
 }}
  \label{fig:confusion_matrix}}
  \vspace{-5mm}
\end{figure*}

For the WST based feature extraction method, averaging each set of scattering coefficients generated from the signal transform resolved to a 1376 feature set. These feature sets are used to training the ML algorithm. Similarly to CWT, we apply PCA to reduce the feature dimensionality with $95\%$ explained variance. The number of principal components (PCs) that accounted for the  $95\%$ explained variance varies depending on the state of the signal. When transforming the transient state of the signal with WST and reducing the feature sets with PCA, it reduces the feature set from 1376 to 131. Conversely, the feature set was reduced from 1376 to 71 using the steady state for feature extraction.

The WST-based feature extraction method exhibits better performance when compared with CWT based feature extraction method. The accuracy of models trained with WST-based coefficients(i.e., scattergram coefficients) for the three-class problem (i.e., group-device classification) is $99.8\%$, $99.5\%$, and $99.9\%$ for \textit{k}NN, SVM, and ensemble, respectively, when utilizing the transient state of the signal for feature extraction. Applying PCA to scattering coefficients of the transient state of the signals does not affect the ML classifiers negatively as compared to CWT coefficients. \textit{k}NN and SVM classification accuracies increased by $0.1\%$ after the scattering coefficients reduction by PCA when exploiting the transient of the signals. On the other hand, the ensemble classifier's average accuracy is reduced by $1.1\%$. More so, using the scattering coefficients of the steady state outperform using the CWT coefficients based on the average classification accuracy of the ML algorithms and SqueezeNet.

Furthermore, using WST-based features for specific-devices classification (i.e., 10 classes), the performance of the ML algorithms and SqueezeNet for both transient and steady state outperforms CWT-based features. For instance, the average accuracy for \textit{k}NN, SVM, ensemble, and SqueezeNet when exploiting the transient of the signals are $88\%$, $88.4\%$, $91\%$, and $88.6\%$, respectively.  More so, there is no significant negative impact on the performance of the ML classifiers when PCA is used to reduce the dimensionality of scattergram coefficients. For example, the performance of \textit{k}NN and SVM classifiers improves after feature reduction to $88.5\%$ and $89.5\%$ respectively using the transient state.

We observed that the classification accuracy of group-device classifiers is higher than the specific-devices classifiers. This is because the misclassification rate is higher among UAVs from the same manufacturer (i.e., the DJI UAVs). Fig.~\ref{fig:confusion_matrix}(b) and Fig.~\ref{fig:confusion_matrix}(d) show examples of two confusion matrices for specific-devices classifiers (i.e., SqueezeNet classifier where the transient state and CWT framework are used and Ensemble + PCA classifier where the transient state and WST framework are utilized). In Fig.~\ref{fig:confusion_matrix}(b), the SqueezeNet classifier misclassifies $36\%$ of DJI Inspire signals as DJI Matrice 600 (i.e., DJI M600) and vice-versa. On the other hand, the Beebeerun and 3DR Iris FSTH9X which are UAVs from different manufacturers are  $96\%$  and $99\%$ correctly classified respectively. Similarly, in Fig.~\ref{fig:confusion_matrix}(d), it can be seen that the errors cluster around the DJI UAVs.

\subsection{Performance at Difference SNR}

\begin{figure*}{}
\center{
\begin{subfloat}[]{\includegraphics[scale=0.6]{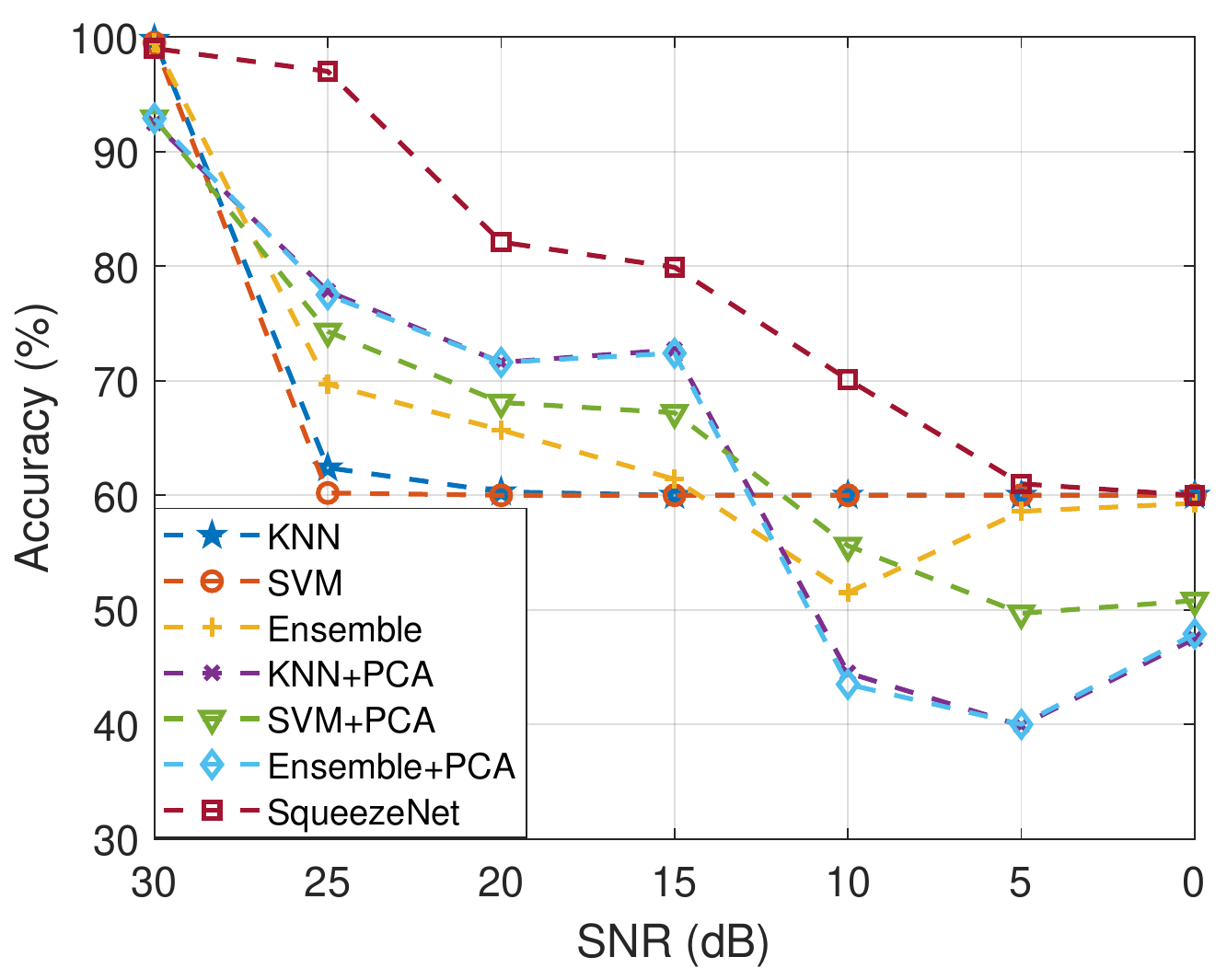}\label{}}
\end{subfloat}
\begin{subfloat}[]{\includegraphics[scale=0.6]{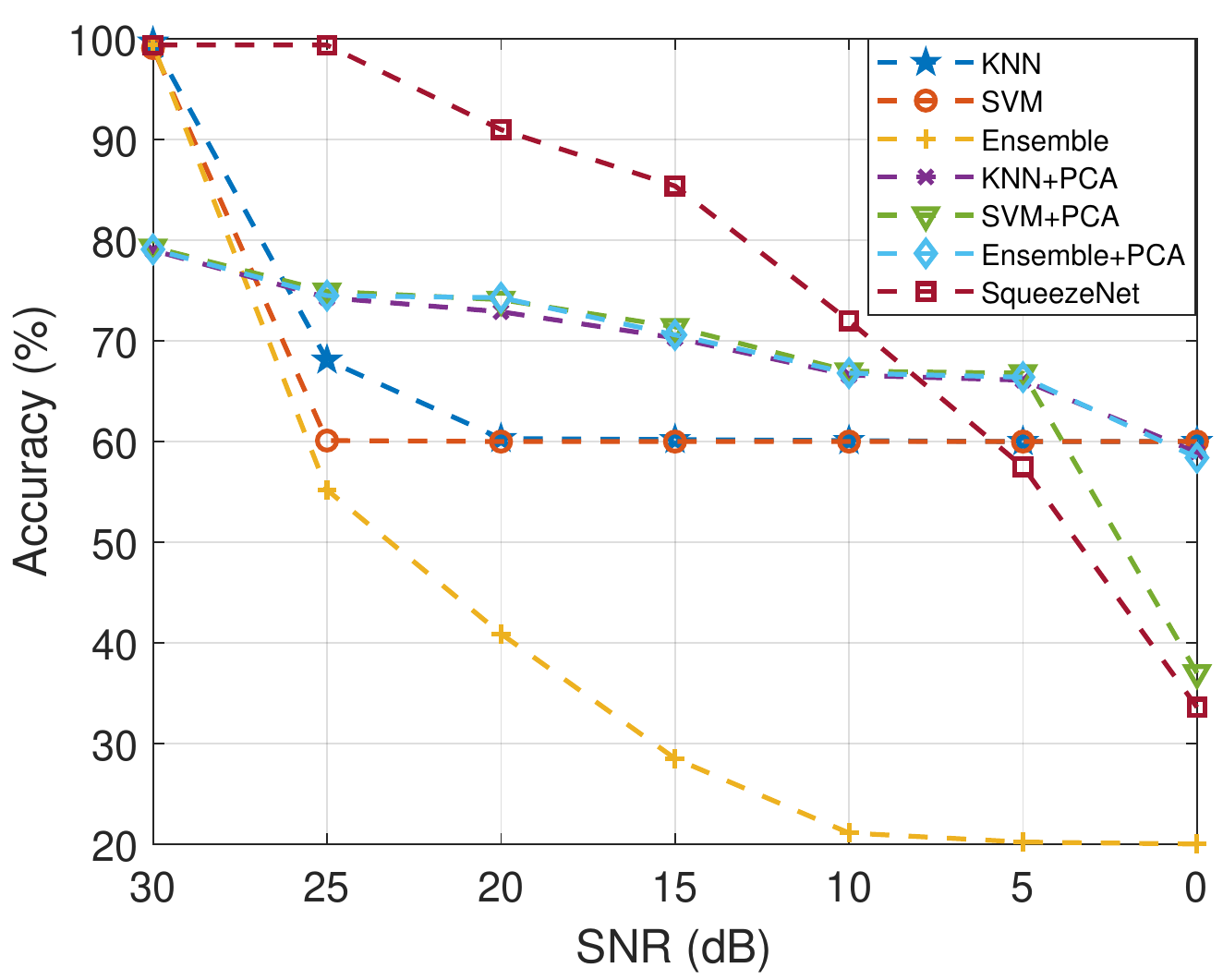}\label{}}
\end{subfloat}
 \begin{subfloat}[]{\includegraphics[scale=0.6]{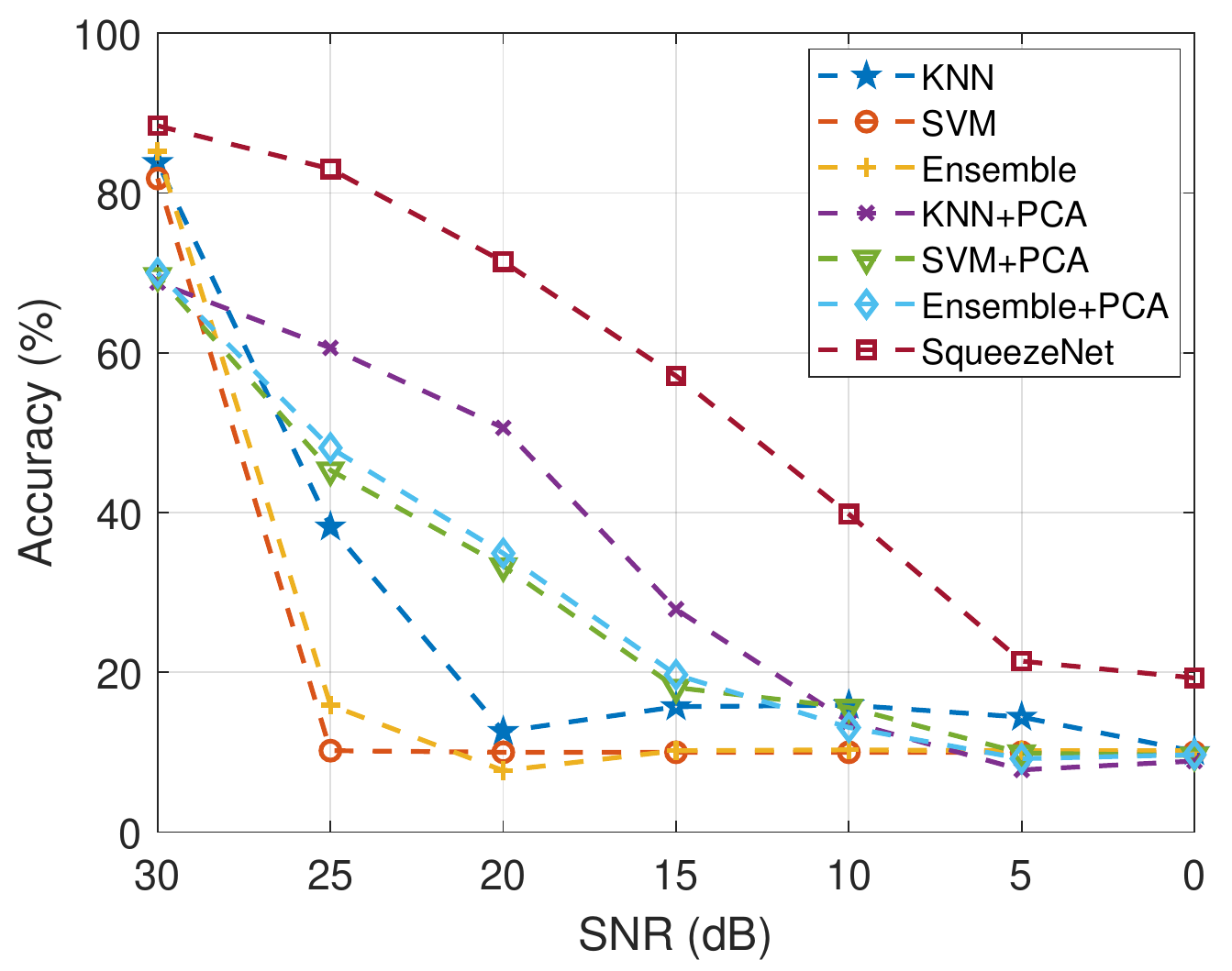}\label{}}
\end{subfloat}
 \begin{subfloat}[]{\includegraphics[scale=0.6]{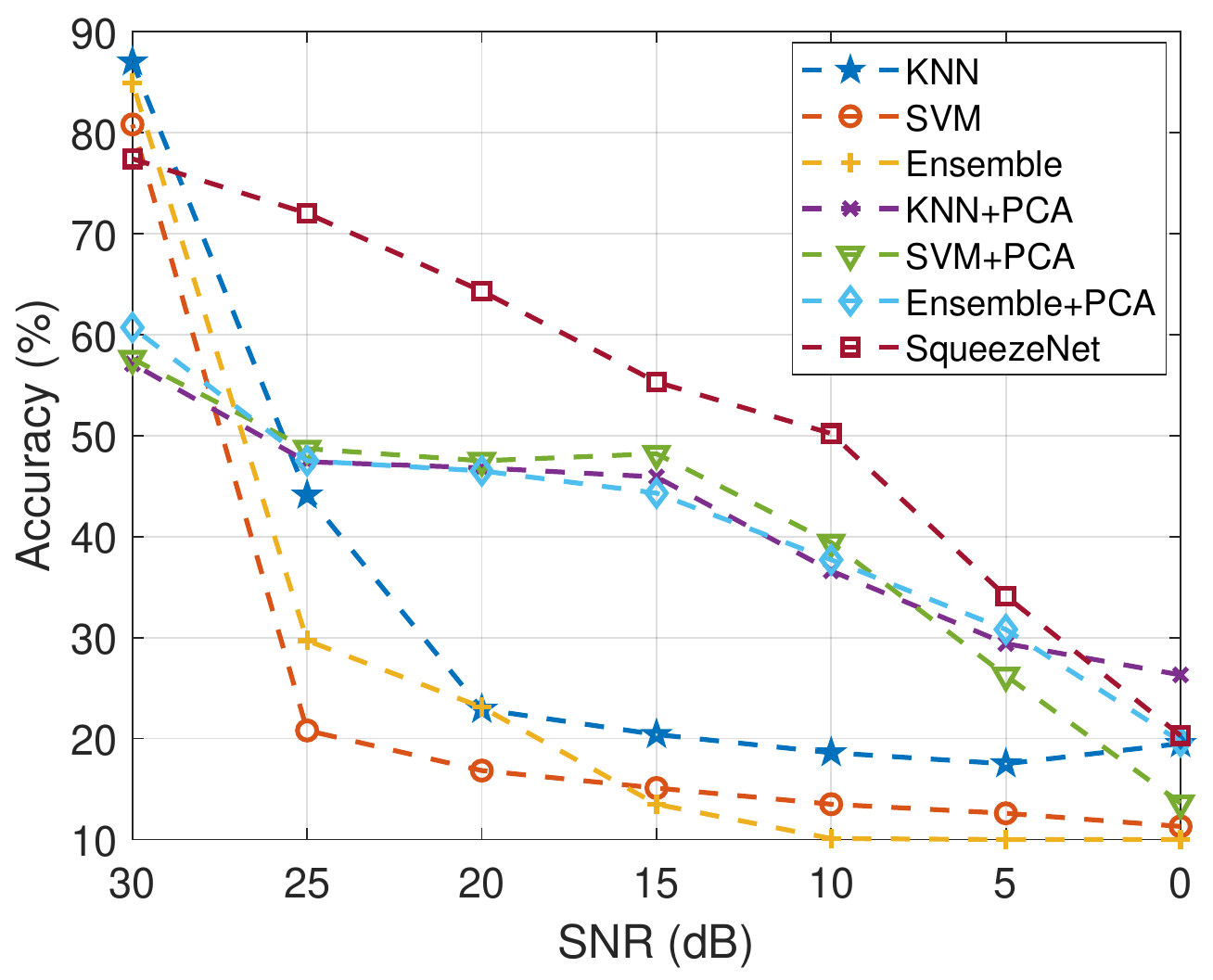}\label{}}
 \end{subfloat}
 \caption{{SNR results: (a) Group-device classification using transient state and CWT framework, (b) Group-device classification using steady state and CWT framework, (c) Specific-device classification using transient state and CWT framework, and (d) Specific-device classification using steady state and CWT framework.}}
  \label{fig:SNR_evalulation_CWT}}
  \vspace{-5mm}
\end{figure*}

\begin{figure*}
\center{

 \begin{subfloat}[]{\includegraphics[scale=0.58]{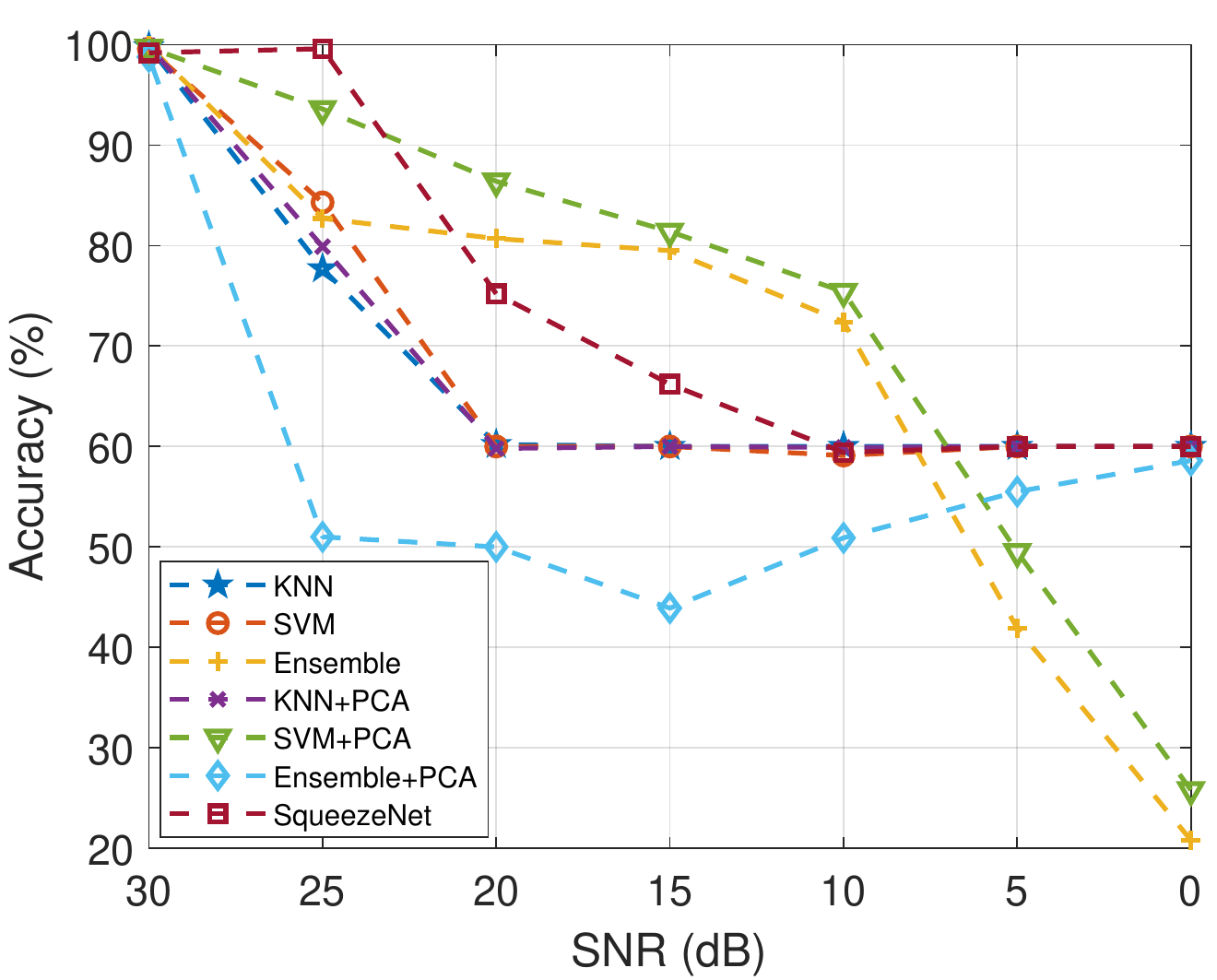}\label{}}
 \end{subfloat}
\begin{subfloat}[]{\includegraphics[scale=0.58]{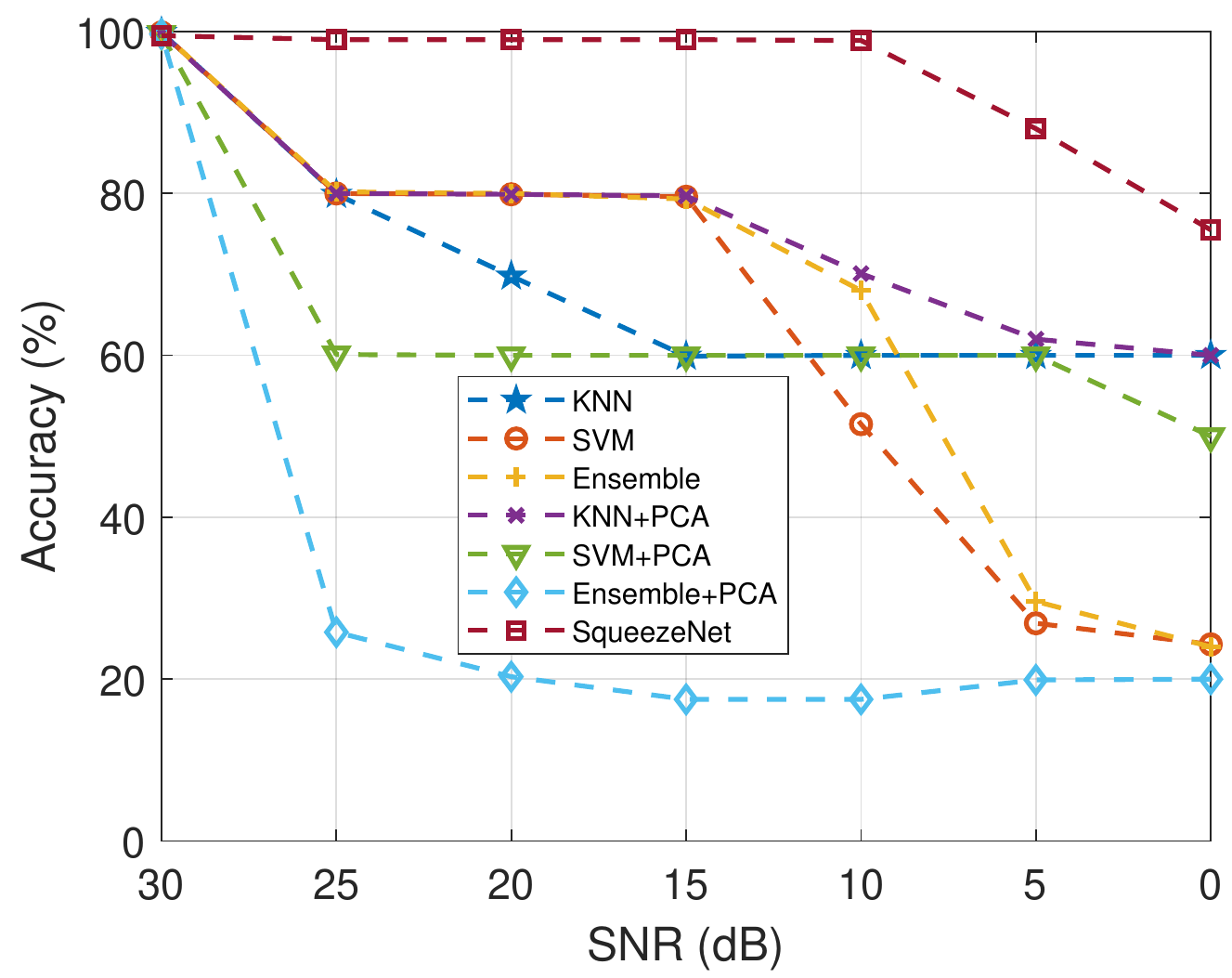}\label{}}
\end{subfloat}

\begin{subfloat}[]{\includegraphics[scale=0.58]{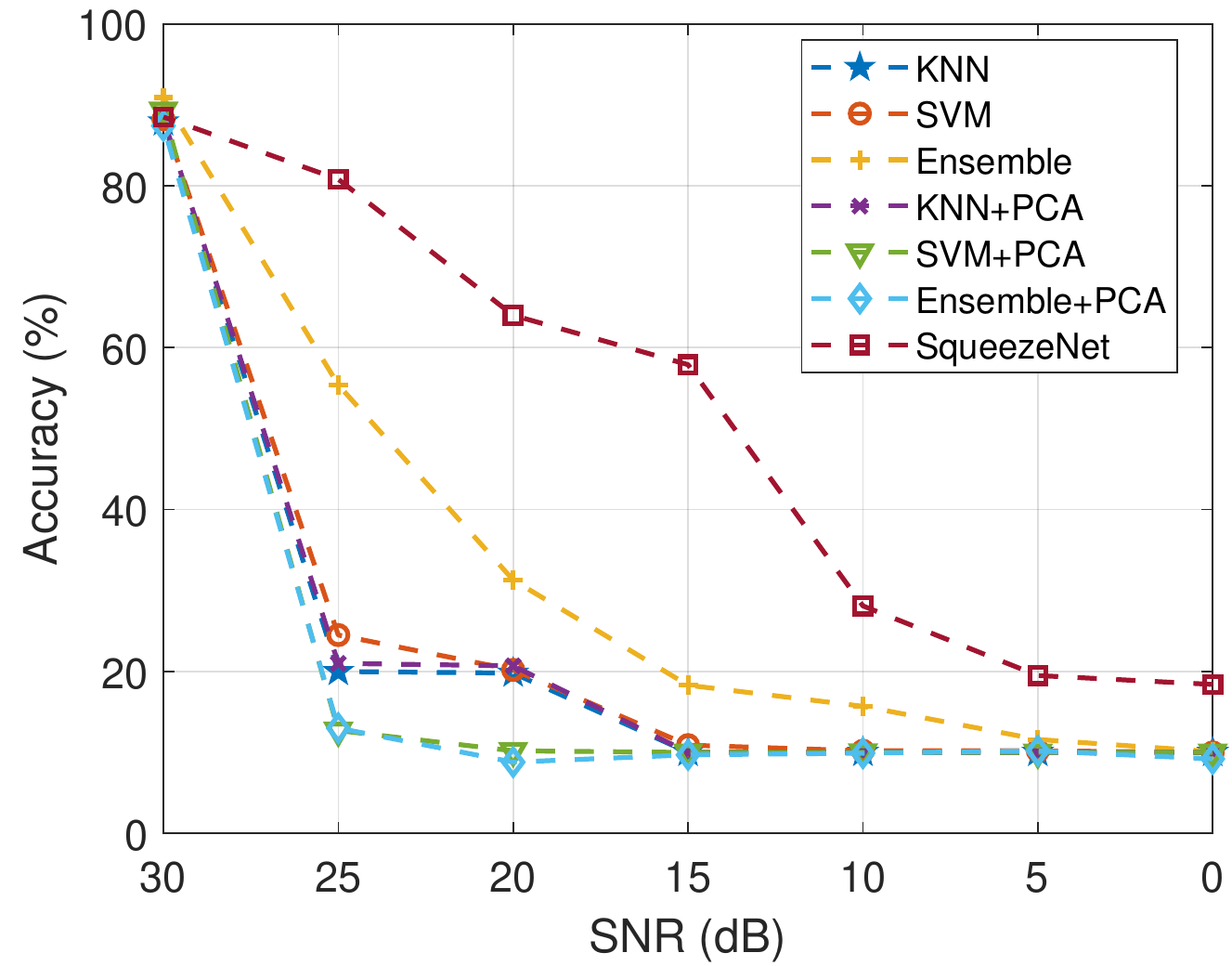}\label{}}
\end{subfloat}
\begin{subfloat}[]{\includegraphics[scale=0.58]{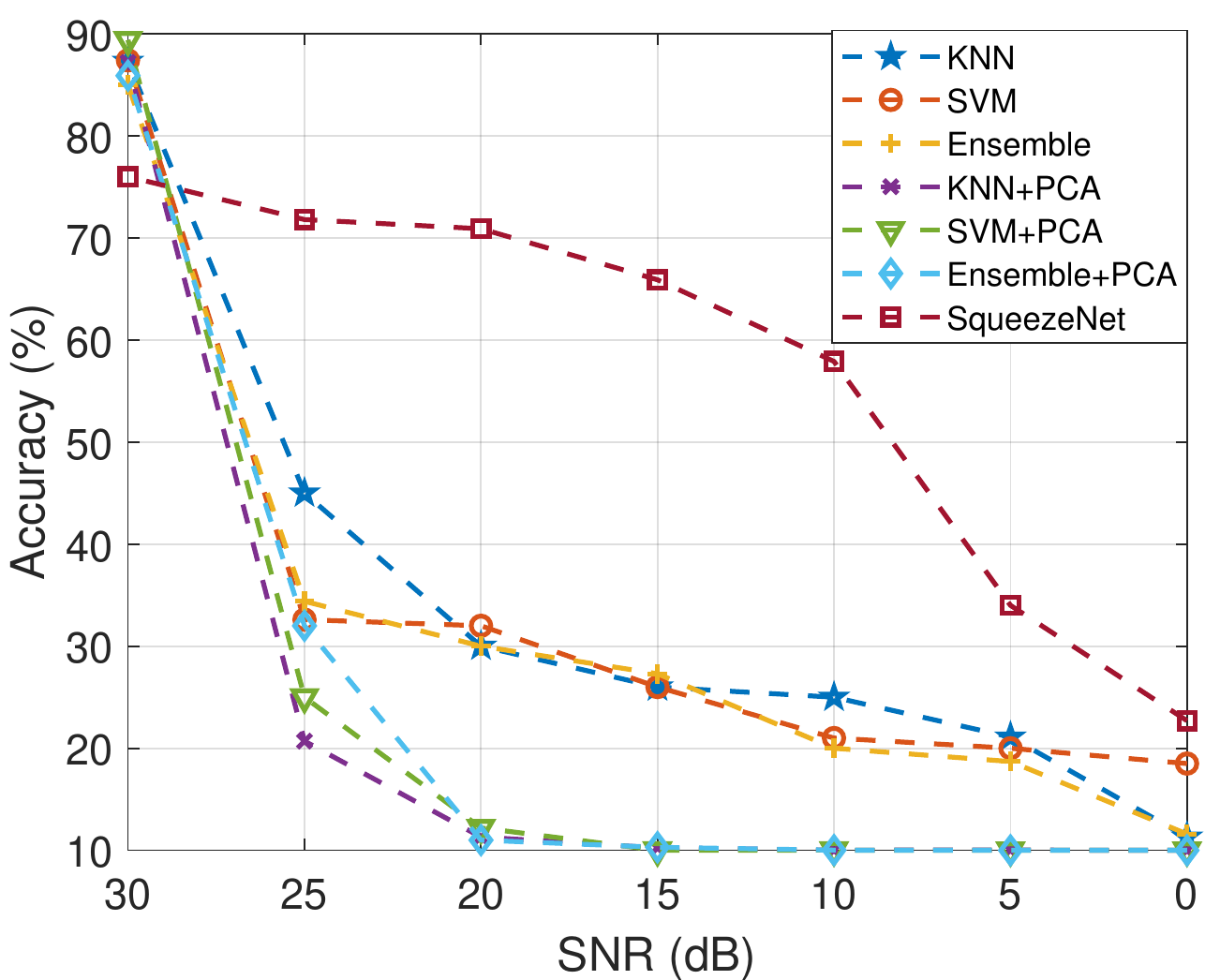}\label{}}
\end{subfloat}
 \caption{{SNR results:  (a) Group-device classification using transient state and WST  framework, (b) Group-device classification using steady state and WST  framework, (c) Specific-device classification using transient state and WST  framework, and  (d) Specific-device classification using steady state and WST framework. }}
  \label{fig:SNR_evalulation_WST}}
  \vspace{-5mm}
\end{figure*}

Due to the nature of RF signals, accuracy or other performance metrics of a classifier is not sufficient in evaluating the proposed classifiers.  The operation environment of an RF device affects the SNR of an RF signal. Because we built our classifiers using signals captured at 30~dB SNR, it is important to evaluate the performance of the classifiers at different SNR levels.  Adding Additive White Gaussian Noise (AWGN) to signals is a common mechanism for varying the SNR of a signal. By varying the SNR of the signals, the behavior of our classifiers is studied.
Fig.~\ref{fig:SNR_evalulation_CWT} and Fig.~\ref{fig:SNR_evalulation_WST} show the performance of our classifiers under varying SNR based on the feature extraction methods (i.e., CWT and WST), the state of signal used for the extraction of features, and the number of classes, respectively.

Fig.~\ref{fig:SNR_evalulation_CWT}(a) shows the performance of the group-device classifiers when CWT is used for extracting features from the transient state of the RF signals. One important factor to note about the test data for group-device classifiers is that it is unbalanced data. That is, $60\%$ of the test data includes UAV signals, while Bluetooth and WiFi signals constitute $20\%$ each. In some cases, when the SNR is below 30 dB and the average accuracy of a group-device classifier is $60\%$, the model classifies all signals as a UAV controller signals. In this instance, the precision and recall of the model are $33.3\%$ and $20\%$, respectively. For example, in Fig.~\ref{fig:SNR_evalulation_CWT}(a), SVM gives an accuracy of $60\%$ from 25~dB to 0~dB. In this instance, any signal that goes through the model is classified as a UAV controller signal. Despite that, the accuracy of the model is $99.5\%$ at the 30~dB SNR. When this case occurs we call it a random classification. All the classifiers with coefficient based signatures
are equal to or below the threshold of 60~ accuracy from an SNR 5 dB and below, as shown in in Fig.~\ref{fig:SNR_evalulation_CWT}(a).

Conversely, in exploiting the steady state of RF signals using CWT for feature extraction under the group-device classification as shown in Fig.~\ref{fig:SNR_evalulation_CWT}(b)., ensemble + PCA,  \textit{k}NN + PCA, and SVM + PCA make non-random classification from 30~dB to 5~dB and there is a steady decrement in the accuracy as the SNR decreases. Similarly, Squeezenet which uses scalogram joined the progression of non-random classification at 10~dB, and the performance of SqueezeNet, in this case, decreases with a decrease in the SNR levels. SqueezeNet outperforms other classifiers at 30~dB to 10~dB SNR as shown in Fig.~\ref{fig:SNR_evalulation_CWT}(a) and Fig.~\ref{fig:SNR_evalulation_CWT}(b).

Fig.~\ref{fig:SNR_evalulation_CWT}(c) shows the performance of specific-device classifiers under varying SNR when CWT is used for signature extraction on the transient state of the signals. Squeezenet outperforms other ML algorithms in terms of responsiveness to varying the SNR. From 0~dB SNR and above, SqueezeNet has higher accuracy compared to other ML algorithms. Similar performance is observed for SqueezeNet from 25~dB to 5~dB when using CWT on the steady state of the signal for specific-device classifiers shown in Fig.~\ref{fig:SNR_evalulation_CWT}(d) Also, in  Fig.~\ref{fig:SNR_evalulation_CWT}(c) and Fig.~\ref{fig:SNR_evalulation_CWT}(d), it can be seen that the ML algorithms and SqueezeNet accuracies are decreasing with decrease in the SNR.

Fig.~\ref{fig:SNR_evalulation_WST}(b) shows the performance of the group-device classifiers where WST is used for extracting features from the steady state of the RF signals. SqueezeNet gives an accuracy that is above $60\%$ at an SNR of 0~dB and this continues to increase as the SNR increases. At 10~dB, the accuracy stands as high as $98.9\%$. This outperforms the scenario when WST is used on the transient state of the signal or CWT for feature extraction as shown in Fig.~\ref{fig:SNR_evalulation_CWT}(a) and Fig.~\ref{fig:SNR_evalulation_CWT}(b) for group device classification.

Fig.~\ref{fig:SNR_evalulation_WST}(c) and Fig.~\ref{fig:SNR_evalulation_WST}(d) depict a specific-device classifier where WST is used for feature extraction on the transient and steady state of the signal, respectively. SqueezeNet also outperforms the ML algorithms from 25~dB to 0~dB.

\subsection{Computational Complexity and Actual Time Cost}
\begin{figure}
\center{\includegraphics[scale=0.55]{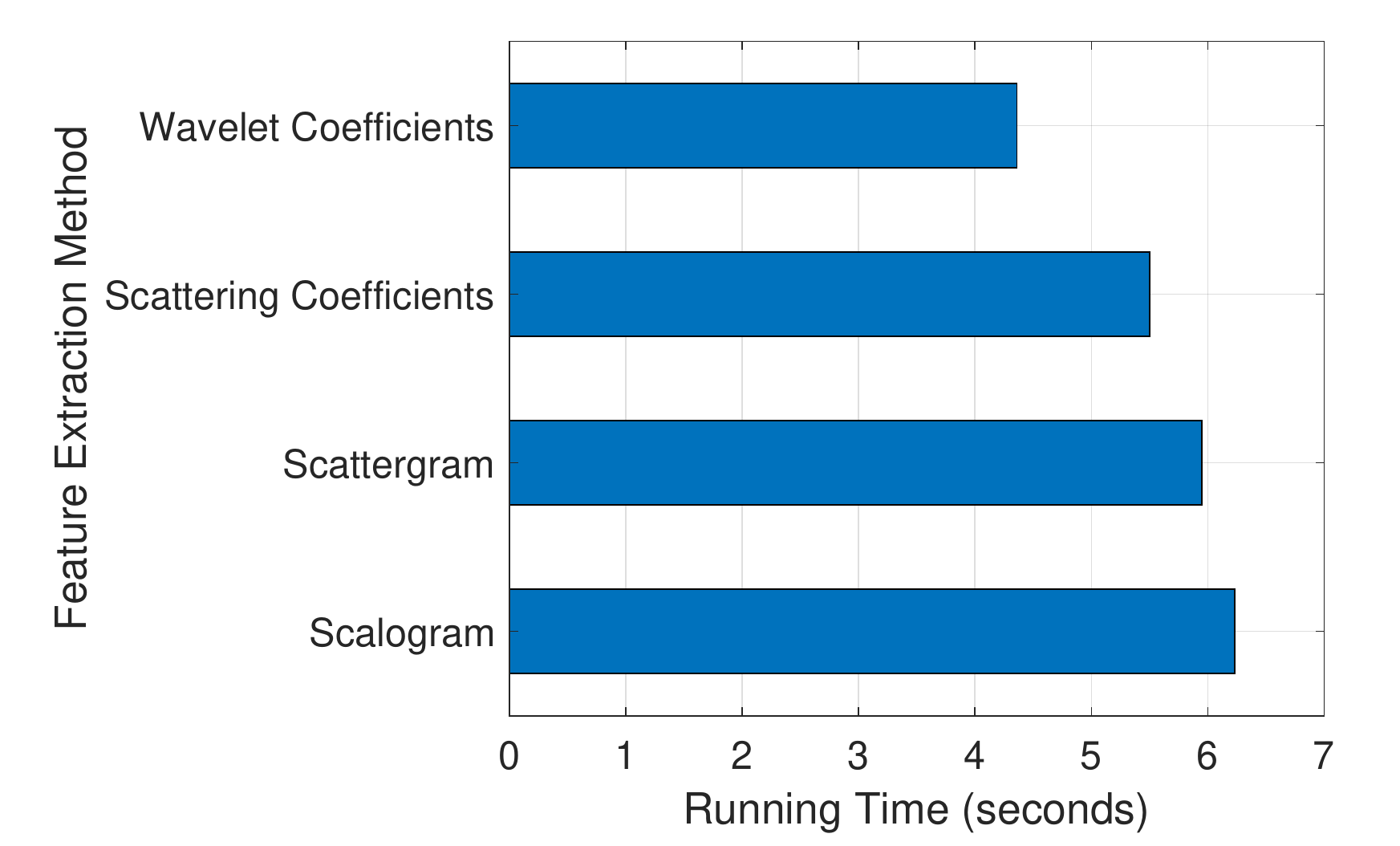}}
\caption{Running time for the feature extraction methods.}
\label{Fig:running_cost_of_feature}
 \vspace{-5mm}
\end{figure}

We considered the computational running time for each feature extraction method and the inference time for each classifier. Fig.~\ref{Fig:running_cost_of_feature} shows the average running time for extracting each category of features. Extracting the image-based features (scalogram and scattergram) from RF signals have higher average running times when compared with the coefficient-based features (wavelet and scattering coefficients). It can be seen in Fig.~\ref{Fig:running_cost_of_feature} that the two image-based features (i.e., scalogram and scattergram) have the highest time cost. Extracting scalogram from the CWT framework has the highest running time. Scattergram has the second-highest running time and it is immediately followed by the scattering coefficient-based method. On the other hand, the wavelet coefficients from the CWT framework has the lowest running time.

Depending on the type of feature used by a classifier, the end-to-end computational time, from the point a signal is captured to inference, is the sum of run time for extracting feature and inference time. Table \ref{Table_inference_time} shows the inference time of the classifiers. From Table \ref{Table_inference_time}, the average inference time of the scattering coefficient-based (i.e., using the coefficient of WST as input features) classifiers is higher than the classifiers that take in CWT coefficients as input features. For instance, average inference time for \textit{k}NN, SVM, and ensemble classifiers under group-device classification using the transient state (i.e., three classes where CWT coefficients are the input feature) are $108$ millisecond, $55.3$ millisecond, and $56.8$ millisecond, respectively. On the other hand, using scattering coefficient-based features from the transient state, the average inference time for \textit{k}NN, SVM, and ensemble classifiers under group-device classification are $822$ millisecond, $155$ millisecond, and $129$ millisecond, respectively. This is primarily due to the higher dimensionality of the scattering coefficient-based features (i.e., 1376 feature set) as compared to wavelet coefficient-based features (i.e., 114 feature set).

Furthermore, the average inference time of SqueezeNet when using scalogram is higher than when using scattergram. For example, when the scalogram from the transient state of an RF signal is used for group-device classification, the inference time is $200$ millisecond. On the other hand, with scattergram, it takes $150$ millisecond.

Overall, exploiting wavelet coefficients for an RF-based UAV classification has a lower time cost over scattering coefficients. Also, exploiting scalogram has a higher time cost over scattergram.

\begin{table*}
\setlength{\tabcolsep}{1.3pt}
\centering

\caption{Computational time cost for classifier's inference given in seconds.}
\label{Table_inference_time}
\begin{tabular}{|c|c|c|c|c|c|c|c|c|c|}

\hline
 s/n & Algorithm & \multicolumn{4}{c|}{Continuous Wavelet Transform}  &  \multicolumn{4}{c|}{Wavelet Scattering Transform} \\   \cline{3-10}

 & & \multicolumn{2}{c|} {3 Classes} &  \multicolumn{2}{c|} {10 Classes} &  \multicolumn{2}{c|} {3 Classes} &  \multicolumn{2}{c|} {10 Classes} \\    \cline{3-10}

& & Transient & Steady & Transient & Steady & Transient & Steady & Transient & Steady\\

 \hline
1 & \textit{k}NN & 0.108 &	0.08  &	0.0873	 & 0.075  &	0.822 &	0.771 &	0.726 &	0.81	 \\

\hline
2 & SVM &  0.0553 &	0.0243	& 0.109	& 0.083 & 	0.155 &	0.117  &	0.423 &	0.405	 \\

\hline
3 & Ensemble & 0.0568  &	0.357  &	0.379  &	0.094  &	0.129 &	0.101 &	1.48 &	0.108	 	 \\

\hline
4 & \textit{k}NN + PCA & 0.061 &	0.107 &	0.0277 &	0.221 &	0.168  &	0.097 &	0.188 &	0.097	 	 \\

\hline
5 & SVM + PCA & 0.081 &	0.0263 &	0.11  &	0.061  &	0.167  &	0.075 &	1.027  &	0.082		 \\

\hline
6 & Ensemble + PCA & 0.187 &	0.7324 &	0.046 &	0.063 &	0.778 &	0.124 &	1.18 &	0.289		 \\

\hline
7 &  SqueezeNet & 0.2	 & 0.18 &	0.192	 & 0.19  &	0.15 & 	0.151 &	0.16 &	0.159	 	 \\

 \hline
\end{tabular}
 \vspace{-3mm}
\end{table*}

\section{Conclusion} \label{eight}
In this paper, we introduced an approach for the detection and identification of UAVs by exploiting the RF signals transmitted by the UAV controller. The communication between a UAV and its flight controller operates at the 2.4 GHz frequency which is the same frequency that WiFi and Bluetooth devices operate. We acquired RF signals from six UAV controllers, two WiFi routers, and two Bluetooth devices. We extracted RF fingerprints from these signals by proposing four methods for extracting a signature from the two-states of the RF signals (transient and steady state of the signal) using wavelet transform analytics (CWT and WST) to differentiate UAV signals from WiFi and Bluetooth signals, and even identify UAV types.

From CWT, two approaches are proposed for feature extraction (wavelet coefficient based and the other is image-based (i.e., scalogram)). Likewise, in WST, we come up with two methods for feature extraction (scattering coefficient based and image-based (scattergram)). Comparing each method, we realized that using an image-based signature to train a pre-trained CNN based model (SqueezeNet) outperforms coefficient based classifiers where classical ML algorithms are used. We obtained an accuracy of $98.9\%$ at 10~dB using the scattergram of RF signals.

In the results, we observed that using wavelet or scattering coefficient for feature extraction is sensitive to variation in SNR because some classifiers failed outrightly when the test signals are at an SNR different from the training signals despite having high accuracy at 30~dB.

Finally, we showed that both transient and steady state of an RF signal carries unique attributes that can be exploited for UAV identification or classification. It is observed that using wavelet transform analytics for the extraction of RF fingerprints on the steady state of RF signals can tolerate varying SNR compared to using transient state when we evaluated the performance of our classifiers under AWGN condition.

In the future, we will be exploring the possibility of using data mining approaches rather than using signal processing approaches such as wavelet analytics for extracting signatures for UAV identification. We want to have a classifier that is resilient to a low level of SNR.

\bibliographystyle{elsarticle-num}
\bibliography{reference.bib}

\begin{thebibliography}{10}
\expandafter\ifx\csname url\endcsname\relax
  \def\url#1{\texttt{#1}}\fi
\expandafter\ifx\csname urlprefix\endcsname\relax\def\urlprefix{URL }\fi
\expandafter\ifx\csname href\endcsname\relax
  \def\href#1#2{#2} \def\path#1{#1}\fi

\bibitem{alsalam2017autonomous}
B.~H.~Y. Alsalam, K.~Morton, D.~Campbell, F.~Gonzalez, Autonomous {UAV} with
  vision based on-board decision making for remote sensing and precision
  agriculture, in: Proc. IEEE Aerosp. Conf., Big Sky, MT, USA, 2017, pp. 1--12.
\newblock \href {https://doi.org/10.1109/AERO.2017.7943593}
  {\path{doi:10.1109/AERO.2017.7943593}}.

\bibitem{banaszek2017application}
A.~Banaszek, A.~Zarnowski, A.~Cellmer, S.~Banaszek, Application of new
  technology data acquisition using aerial {UAV} digital images for the needs
  of urban revitalization, in: Proc. Environmental Engineering, Int. Conf.,
  Vilnius, Lithuania, 2017, pp. 1--7.
\newblock \href {https://doi.org/10.3846/enviro.2017.159}
  {\path{doi:10.3846/enviro.2017.159}}.

\bibitem{coveney2017lightweight}
S.~Coveney, K.~Roberts, Lightweight {UAV} digital elevation models and
  orthoimagery for environmental applications: data accuracy evaluation and
  potential for river flood risk modelling, Int. Journal of Remote Sensing
  38~(8-10) (2017) 3159--3180.

\bibitem{uasbythenumbers_2020}
FAA, \href{https://www.faa.gov/uas/resources/by_the_numbers/}{U{A}{S} by the
  numbers}, "Accessed: 2020-10-12" (2020).
\newline\urlprefix\url{https://www.faa.gov/uas/resources/by_the_numbers/}

\bibitem{rattledrone2016}
M.~S. Schmidt, M.~D. Shear, \href{shorturl.at/eENOZ}{A drone, too small for
  radar to detect, rattles the white house}, accessed: 2020-10-12 (2016).
\newline\urlprefix\url{shorturl.at/eENOZ}

\bibitem{smuggler_drone2018}
G.~Harkins, \href{shorturl.at/moqG1}{Illicit drone flights surge along
  {U.S.-Mexico} border as smugglers hunt for soft spots}, accessed: 2020-10-12.
\newline\urlprefix\url{shorturl.at/moqG1}

\bibitem{faa_uas_sight_reporting2020}
FAA,
  \href{https://www.faa.gov/uas/resources/public_records/uas_sightings_report/}{U{A}{S}
  sightings report}, accessed: 2020-10-12 (2020).
\newline\urlprefix\url{https://www.faa.gov/uas/resources/public_records/uas_sightings_report/}

\bibitem{birch2015uas}
G.~C. Birch, J.~C. Griffin, M.~K. Erdman, U{A}{S} detection, classification,
  and neutralization: Market survey 2015, Sandia National Laboratories (2015).

\bibitem{doroftei2018qualitative}
D.~Doroftei, G.~De~Cubber, Qualitative and quantitative validation of drone
  detection systems, in: Proc. Int. Symposium on Measurement and Control in
  Robotics, no. 21st, Mons, Belgium, 2018, pp. 26--28.

\bibitem{sturdivant2017systems}
R.~L. Sturdivant, E.~K. Chong, Systems engineering baseline concept of a
  multispectral drone detection solution for airports, IEEE Access 5 (2017)
  7123--7138.
\newblock \href {https://doi.org/10.1109/ACCESS.2017.2697979}
  {\path{doi:10.1109/ACCESS.2017.2697979}}.

\bibitem{brown2017pondering}
B.~Brown, K.~Buckler, Pondering personal privacy: a pragmatic approach to the
  fourth amendment protection of privacy in the information age, Contemporary
  Justice Review 20~(2) (2017) 227--254.

\bibitem{nguyen2018cost}
P.~Nguyen, H.~Truong, M.~Ravindranathan, A.~Nguyen, R.~Han, T.~Vu,
  Cost-effective and passive {RF}-based drone presence detection and
  characterization, GetMobile: Mobile Computing and Commun. 21~(4) (2018)
  30--34.

\bibitem{mezei2015drone}
J.~Mezei, V.~Fiaska, A.~Moln{\'a}r, Drone sound detection, in: Proc. IEEE Int.
  Symposium on Computational Intelligence and Inf. (CINTI), Budapest, Hungary,
  2015, pp. 333--338.

\bibitem{mezei2016drone}
J.~Mezei, A.~Moln{\'a}r, Drone sound detection by correlation, in: Proc. IEEE
  Int. Symposium on Applied Computational Intelligence and Inf. (SACI),
  Timisoara, Romania, 2016, pp. 509--518.
\newblock \href {https://doi.org/10.1109/SACI.2016.7507430}
  {\path{doi:10.1109/SACI.2016.7507430}}.

\bibitem{nijim2016drone}
M.~Nijim, N.~Mantrawadi, Drone classification and identification system by
  phenome analysis using data mining techniques, in: Proc. IEEE Symposium on
  Technologies for Homeland Security (HST), Waltham, MA, USA, 2016, pp. 1--5.
\newblock \href {https://doi.org/10.1109/THS.2016.7568949}
  {\path{doi:10.1109/THS.2016.7568949}}.

\bibitem{yue2018software}
X.~Yue, Y.~Liu, J.~Wang, H.~Song, H.~Cao, Software defined radio and wireless
  acoustic networking for amateur drone surveillance, IEEE Commun. Mag. 56~(4)
  (2018) 90--97.

\bibitem{shi2018anti}
X.~Shi, C.~Yang, W.~Xie, C.~Liang, Z.~Shi, J.~Chen, Anti-drone system with
  multiple surveillance technologies: Architecture, implementation, and
  challenges, IEEE Commun. Mag. 56~(4) (2018) 68--74.

\bibitem{saqib2017study}
M.~Saqib, S.~D. Khan, N.~Sharma, M.~Blumenstein, A study on detecting drones
  using deep convolutional neural networks, in: Proc. IEEE Int. Conf. on Adv.
  Video and Signal Based Surveillance (AVSS), Lecce, Italy, 2017, pp. 1--6.
\newblock \href {https://doi.org/10.1109/AVSS.2017.8078558}
  {\path{doi:10.1109/AVSS.2017.8078558}}.

\bibitem{schumann2017deep}
A.~Schumann, L.~Sommer, J.~Klatte, T.~Schuchert, J.~Beyerer, Deep cross-domain
  flying object classification for robust {U}{A}{V} detection, in: Proc. IEEE
  Int. Conf. on Advanced Video and Signal Based Surveillance (AVSS), Lecce,
  Italy, 2017, pp. 1--6.
\newblock \href {https://doi.org/10.1109/AVSS.2017.8078558}
  {\path{doi:10.1109/AVSS.2017.8078558}}.

\bibitem{unlu2019deep}
E.~Unlu, E.~Zenou, N.~Riviere, P.-E. Dupouy, Deep learning-based strategies for
  the detection and tracking of drones using several cameras, IPSJ Trans. on
  Comput. Vision and Applications 11~(1) (2019) 7.

\bibitem{andravsi2017night}
P.~Andra{\v{s}}i, T.~Radi{\v{s}}i{\'c}, M.~Mu{\v{s}}tra, J.~Ivo{\v{s}}evi{\'c},
  Night-time detection of {UAV}s using thermal infrared camera, Transportation
  Research Procedia 28 (2017) 183--190.

\bibitem{drozdowicz201635}
J.~Drozdowicz, M.~Wielgo, P.~Samczynski, K.~Kulpa, J.~Krzonkalla, M.~Mordzonek,
  M.~Bryl, Z.~Jakielaszek, 35 {GHz FMCW} drone detection system, in: Proc. Int.
  Radar Symposium (IRS), Krakow, Poland, 2016, pp. 1--4.
\newblock \href {https://doi.org/10.1109/IRS.2016.7497351}
  {\path{doi:10.1109/IRS.2016.7497351}}.

\bibitem{mendis2016deep}
G.~J. Mendis, T.~Randeny, J.~Wei, A.~Madanayake, Deep learning based doppler
  radar for micro {U}{A}{S} detection and classification, in: Proc. IEEE
  Military Commun. Conf., Baltimore, MD, USA, 2016, pp. 924--929.
\newblock \href {https://doi.org/10.1109/MILCOM.2016.7795448}
  {\path{doi:10.1109/MILCOM.2016.7795448}}.

\bibitem{zhao2018classification}
C.~Zhao, C.~Chen, Z.~Cai, M.~Shi, X.~Du, M.~Guizani, Classification of small
  {UAV}s based on auxiliary classifier wasserstein {GAN}s, in: Proc. IEEE
  Global Commun. Conf. (GLOBECOM), Abu Dhabi, United Arab Emirates, 2018, pp.
  206--212.
\newblock \href {https://doi.org/10.1109/GLOCOM.2018.8647973}
  {\path{doi:10.1109/GLOCOM.2018.8647973}}.

\bibitem{zhou2018unmanned}
W.~Zhou, L.~Wang, B.~Lu, N.~Jin, L.~Guo, J.~Liu, H.~Sun, H.~Liu, Unmanned
  aerial vehicle detection based on channel state information, in: IEEE Int.
  Conf. on Sensing, Commun. Networking (SECON Workshops), Hong Kong, China,
  2018, pp. 1--5.
\newblock \href {https://doi.org/10.1109/SECONW.2018.8396360}
  {\path{doi:10.1109/SECONW.2018.8396360}}.

\bibitem{ezuma2019micro}
M.~Ezuma, F.~Erden, C.~K. Anjinappa, O.~Ozdemir, I.~Guvenc, Micro-{UAV}
  detection and classification from {RF} fingerprints using machine learning
  techniques, in: Proc. IEEE Aerosp. Conf., Big Sky, Montana, 2019.

\bibitem{ezuma2019detection}
M.~Ezuma, F.~Erden, C.~K. Anjinappa, O.~Ozdemir, I.~Guvenc, Detection and
  classification of {UAV}s using {RF} fingerprints in the presence of {WiFi}
  and bluetooth interference, IEEE Open Journal of the Commun. Society 1 (2019)
  60--76.
\newblock \href {https://doi.org/10.1109/OJCOMS.2019.2955889}
  {\path{doi:10.1109/OJCOMS.2019.2955889}}.

\bibitem{al2019rf}
M.~F. Al-Sa’d, A.~Al-Ali, A.~Mohamed, T.~Khattab, A.~Erbad, {RF}-based drone
  detection and identification using deep learning approaches: An initiative
  towards a large open source drone database, Future Generation Comput. Syst.
  100 (2019) 86--97.

\bibitem{alipour2019machine}
A.~Alipour-Fanid, M.~Dabaghchian, N.~Wang, P.~Wang, L.~Zhao, K.~Zeng, Machine
  learning-based delay-aware {UAV} detection and operation mode identification
  over encrypted {WiFi} traffic, IEEE Trans. on Inf. Forensics and Security 15
  (2019) 2346--2360.

\bibitem{kennedy2008radio}
I.~O. Kennedy, P.~Scanlon, F.~J. Mullany, M.~M. Buddhikot, K.~E. Nolan, T.~W.
  Rondeau, Radio transmitter fingerprinting: A steady state frequency domain
  approach, in: Proc. IEEE Veh. Technol. Conf., Calgary, BC, Canada, 2008, pp.
  1--5.
\newblock \href {https://doi.org/10.1109/VETECF.2008.291}
  {\path{doi:10.1109/VETECF.2008.291}}.

\bibitem{klein2009application}
R.~W. Klein, M.~A. Temple, M.~J. Mendenhall, Application of wavelet-based {RF}
  fingerprinting to enhance wireless network security, Journal of Commun. and
  Net. 11~(6) (2009) 544--555.

\bibitem{fulcher2014highly}
B.~D. Fulcher, N.~S. Jones, Highly comparative feature-based time-series
  classification, IEEE Trans. on Knowledge and Data Engineering 26~(12) (2014)
  3026--3037.

\bibitem{wang2013experimental}
X.~Wang, A.~Mueen, H.~Ding, G.~Trajcevski, P.~Scheuermann, E.~Keogh,
  Experimental comparison of representation methods and distance measures for
  time series data, Data Mining and Knowledge Discovery 26~(2) (2013) 275--309.

\bibitem{hall2003detection}
J.~Hall, M.~Barbeau, E.~Kranakis, Detection of transient in radio frequency
  fingerprinting using signal phase, Wireless and Optical Commun. (2003)
  13--18.

\bibitem{allahham2020deep}
M.~S. Allahham, T.~Khattab, A.~Mohamed, Deep learning for {RF}-based drone
  detection and identification: A multi-channel 1-d convolutional neural
  networks approach, in: Proc. IEEE Int. Conf. on Inf., IoT, and Enabling
  Technologies (ICIoT), Doha, Qatar, 2020, pp. 112--117.
\newblock \href {https://doi.org/10.1109/ICIoT48696.2020.9089657}
  {\path{doi:10.1109/ICIoT48696.2020.9089657}}.

\bibitem{ozturk2020rf}
E.~Ozturk, F.~Erden, I.~Guvenc, {RF}-based low-{SNR} classification of {UAVs}
  using convolutional neural networks, arXiv preprint arXiv:2009.05519 (2020).

\bibitem{mallat1999wavelet}
S.~Mallat, A wavelet tour of signal processing, Elsevier, 1999.

\bibitem{adisson2002illustrated}
P.~S. Adisson, The illustrated wavelet transform handbook: Introductory theory
  and applications in science, engineering, medicine and finance (2002).

\bibitem{rioul1991wavelets}
O.~Rioul, M.~Vetterli, Wavelets and signal processing, IEEE Signal Processing
  Mag. 8~(4) (1991) 14--38.

\bibitem{gao2010wavelets}
R.~X. Gao, R.~Yan, Wavelets: {T}heory and applications for manufacturing,
  Springer Science \& Business Media, 2010.

\bibitem{anden2014deep}
J.~And{\'e}n, S.~Mallat, Deep scattering spectrum, IEEE Trans.on Signal
  Processing 62~(16) (2014) 4114--4128.

\bibitem{bruna2013invariant}
J.~Bruna, S.~Mallat, Invariant scattering convolution networks, IEEE Trans. on
  Pattern Analysis and Machine Intelligence 35~(8) (2013) 1872--1886.

\bibitem{mallat2012group}
S.~Mallat, Group invariant scattering, Commun. on Pure and Applied Mathematics
  65~(10) (2012) 1331--1398.

\bibitem{mallat2016understanding}
M.~St{\'e}phane, Understanding deep convolutional networks, Philosophical
  Transactions of the Royal Society A: Mathematical, Physical and Engineering
  Sciences 374~(2065) (2016) 20150203.

\bibitem{shen2006review}
L.~Shen, L.~Bai, A review on gabor wavelets for face recognition, Pattern
  analysis and applications 9~(2) (2006) 273--292.

\bibitem{iandola2016squeezenet}
F.~N. Iandola, S.~Han, M.~W. Moskewicz, K.~Ashraf, W.~J. Dally, K.~Keutzer,
  {SqueezeNet}: {AlexNet}-level accuracy with 50x fewer parameters and< 0.5
  {MB} model size, arXiv preprint arXiv:1602.07360 (2016).

\bibitem{pan2009survey}
S.~J. Pan, Q.~Yang, A survey on transfer learning, IEEE Trans. on Knowledge and
  Data Engineering 22~(10) (2009) 1345--1359.

\bibitem{snoek2012practical}
J.~Snoek, H.~Larochelle, R.~P. Adams, Practical bayesian optimization of
  machine learning algorithms, Advances in Neural Inf. Processing syst. 25
  (2012) 2951--2959.

\bibitem{kantardzic2011data}
M.~Kantardzic, Data mining: concepts, models, methods, and algorithms, John
  Wiley \& Sons, 2011.

\end{thebibliography}




\end{document}